\documentclass[11pt,letter]{emulateapj}
\usepackage{graphicx}
\usepackage{amsmath}
\usepackage{amssymb}
\usepackage{hyperref}

\shorttitle{Faraday Depth Survey of the Northern Sky}
\shortauthors{M. Wolleben et al.}

\begin{document}

\title{The Global Magneto-Ionic Medium Survey: A Faraday Depth Survey of the Northern Sky Covering 1280 to 1750~MHz}

\author{M. Wolleben\altaffilmark{1,2}, T.L. Landecker\altaffilmark{1}, K.A.
Douglas\altaffilmark{1,3,4}, A.D. Gray\altaffilmark{1},  A.
Ordog\altaffilmark{3,5,1}, J.M. Dickey\altaffilmark{6}, A.S. Hill\altaffilmark{5,1},
E. Carretti\altaffilmark{7}, J.C. Brown\altaffilmark{3}, B.M.
Gaensler\altaffilmark{8}, J.L. Han\altaffilmark{9,10,11}, M.
Haverkorn\altaffilmark{12}, R. Kothes \altaffilmark{1}, J.P.
Leahy\altaffilmark{13}, N. McClure-Griffiths\altaffilmark{14}, D.
McConnell\altaffilmark{15,14}, W. Reich\altaffilmark{16}, A.R.
Taylor\altaffilmark{17,18}, A.J.M. Thomson\altaffilmark{19}, J.L. West\altaffilmark{8}}

\altaffiltext{1}{National Research Council Canada, Herzberg Research Centre for Astronomy and Astrophysics, Dominion Radio Astrophysical Observatory, P.O. Box 248, Penticton, British Columbia, V2A 6J9, Canada}

\altaffiltext{2}{Skaha Remote Sensing Ltd., 3165 Juniper Drive, Naramata,
British Columbia V0H 1N0, Canada}

\altaffiltext{3}{Department of Physics and Astronomy, University of
Calgary, 2500 University Drive, Calgary, AB, T2N 1N4, Canada}

\altaffiltext{4}{Physics and Astronomy Department, Okanagan College,
1000 KLO Road, Kelowna, British Columbia, V1Y 4X8, Canada}

\altaffiltext{5}{Department of Computer Science, Mathematics, Physics, and Statistics, University of British Columbia, Okanagan Campus, 3187 University Way, Kelowna, British Columbia, V1V 1V7, Canada}

\altaffiltext{6}{School of Natural Sciences, Private Bag 37, University of
Tasmania, Hobart, Tasmania 7001, Australia}

\altaffiltext{7}{INAF - Istituto di Radioastronomia, via P. Gobetti 101, I-40129 Bologna, Italy}

\altaffiltext{8}{Dunlap Institute for Astronomy and Astrophysics, University of Toronto, 50 St.\,George Street, Toronto, M5S 3H4, Canada}

\altaffiltext{9}{National Astronomical Observatories, CAS, Jia-20
DaTun Road, Chaoyang District, Beijing 100101, People's Republic of China}
\altaffiltext{10}{School of Astronomy and Space Sciences, University of the Chinese Academy of Sciences, Beijing, 100049, People's Republic of China}

\altaffiltext{11}{CAS Key Laboratory of FAST, NAOC, Chinese Academy of Sciences, Beijing 100101, People's Republic of China}

\altaffiltext{12}{Radboud University Nijmegen, PO Box 9010, 6500 GL
Nijmegen, the Netherlands}

\altaffiltext{13}{Jodrell Bank Centre for Astrophysics, Department of
Physics and Astronomy, The University of Manchester, Manchester M13 9PL, UK}

\altaffiltext{14}{Research School of Astronomy and Astrophysics, Australian
National University, Cotter Road, Weston Creek, ACT 2611, Australia}

\altaffiltext{15}{CSIRO Astronomy \& Space Science, P.O. Box 76, Epping, New South Wales 1710, Australia}

\altaffiltext{16}{Max-Planck-Institut f\"ur Radioastronomie, 53121
Bonn, Auf dem H\"ugel 69, Germany}

\altaffiltext{17}{Department of Astronomy, University of Cape Town, Rondenbosch 7701, Republic of South Africa}

\altaffiltext{18}{Department of Physics, University of Western Cape, Republic of South Africa}

\altaffiltext{19}{CSIRO Astronomy \& Space Science, PO Box 1130, Bentley, Western Australia 6102, Australia}


\begin{abstract}
{The Galactic interstellar medium hosts a significant magnetic field, which can
be probed through the synchrotron emission produced from its interaction with
relativistic electrons. Linearly polarized synchrotron emission is generated
throughout the Galaxy and, at longer wavelengths, modified along nearly every
path by Faraday rotation in the intervening magneto-ionic medium. Full
characterization of the polarized emission requires wideband observations with
many frequency channels.  We have surveyed polarized radio emission from the
Northern sky over the range 1280 to 1750\,MHz, with channel width 236.8\,kHz,
using the John A. Galt Telescope (diameter 25.6\,m) at the Dominion Radio
Astrophysical Observatory, as part of the Global Magneto-Ionic Medium Survey
(GMIMS). The survey covered 72\% of the sky, declinations $-30^{\circ}$ to
$+87^{\circ}$ at all right ascensions. The intensity scale was absolutely
calibrated, based on the flux density and spectral index of Cygnus\,A.
Polarization angle was calibrated using the extended polarized emission of the
Fan Region. Data are presented as brightness temperatures with angular
resolution $40'$. Sensitivity in Stokes $Q$ and $U$ is 45\,mK rms in a 1.18\,MHz
band. We have applied Rotation Measure Synthesis to the data to obtain a Faraday
depth cube of resolution 150\,${\rm{rad}}\thinspace{\rm{m}}^{-2}$ and
sensitivity 3\,mK rms of polarized intensity. Features in Faraday depth up to a
width of 110\,${\rm{rad}}\thinspace{\rm{m}}^{-2}$ are represented. The maximum
detectable Faraday depth is
${\pm}2{\times}10^4\,{\rm{rad}}\thinspace{\rm{m}}^{-2}$. The survey data are
available at the Canadian Astronomy Data Centre.}
\end{abstract}

\keywords{ISM: magnetic fields, polarization, radio continuum: ISM,
surveys, techniques: polarimetric}

\section{Introduction}
\label{intro}

The magnetic field of the Galaxy is a significant reservoir of energy within the
interstellar medium \citep{ferr01,heil12}. It supports the Galactic disk
\citep{boul90,hill12}, it is profoundly influential in star formation
\citep{pado11}, and it is central to particle acceleration \citep{urov19}.
Theories have been developed of the origin of the field in a Galactic dynamo
\citep{beck96,moss19} and of the impact of the magnetic energy reservoir in
shaping galaxies \citep{kim96}. While its significance is well appreciated \citep{han17}, the magnetic field remains a component of the interstellar medium that is difficult to observe and measure.

Of interest to us is synchrotron emission, generated throughout the Galaxy when
relativistic electrons interact with Galactic magnetic fields. The magnetic
field imprints its direction on the radio signal, which is linearly polarized
with orientation perpendicular to the field at the point of emission. At short
wavelengths, synchrotron emission carries its polarization state to our
telescopes (for example, the WMAP data at 23\,GHz -- \citealp{benn13}) and
yields a two-dimensional portrait of the magnetic field configuration in the
Galaxy. At longer radio wavelengths the polarization state is altered, often
profoundly, by Faraday rotation occurring in magnetized ionized regions along
the propagation path. Synchrotron emission is generated throughout the Galaxy
and Faraday rotation occurs everywhere; the consequent intermingling of emission
and rotation complicates interpretation of polarization observations. Faraday
rotation largely obscures the original field directions; nevertheless, it can be
exploited to give three-dimensional information on magnetic field configurations
in the intervening medium.

Extensive surveys at single frequencies (e.g.
\citealp{brou76,reic04,woll06,sun07,test08,gao10,carr19}) and aperture-synthesis
surveys in the Galactic plane (\citealp{have06a,land10}) have provided
two-dimensional portraits of the polarized radio sky. In combination with
Faraday rotation towards point sources (\citealp{han06,brow07,tayl09,vane21}) these surveys have contributed to
three-dimensional reconstructions of the magnetic field in the Galactic disk and
halo \citep{sun08,vane11,jans12,jaff13,jaff19}. Here we take the next step,
mapping Faraday depth over the entire sky to generate a dataset that can further
elucidate the three-dimensional structure of the Galactic magnetic field.

A source of polarized radio emission is described by the complex
polarization vector at the point of emission,
\begin{equation} 
{{\bf{P}}_0} = Q + iU = {P_0}{e^{2i{\chi_{o}}}}, 
\end{equation} 
where $Q$ and $U$ are the Stokes parameters describing the state of linear
polarization, $P_0$ is the polarized intensity, and $\chi_{o}$ is the polarization
angle. If a Faraday rotating region, entirely separate from the emission region, lies along the intervening path, then, at wavelength $\lambda$, the polarization angle is rotated by 
\begin{equation}
{\Delta}{\chi}={{0.812}{\lambda^2}{\thinspace}{\int}{n_e}{\thinspace}{B_{||}}{\thinspace}dl}={\lambda^2}{\thinspace}{\rm{RM}},
\label{delta_chi}
\end{equation}
where  ${B_{||}}$ is the line-of-sight component of the magnetic
field in $\mu$G, ${n_e}$ is the electron density in cm$^{-3}$, $l$ is the path
length in parsecs, and the integral is computed along the entire line of sight through the Faraday rotating region from the source to the observer. After Faraday rotation the observed polarization vector is 
\begin{equation}
{{\bf{P}}({\lambda}^2)} = {P_0}{e^{2i({{\chi_{o}}}+{\lambda}^{2}{\thinspace}{\rm{RM}})}}
={{\bf{P}}_0}{e^{2i{\lambda}^{2}{\thinspace}{\rm{RM}}}}. 
\label{rotated_P}
\end{equation}
${\rm{RM}}$ in Equations~\ref{delta_chi} and \ref{rotated_P} is the Rotation Measure, a characteristic of the Faraday rotating region, which can be measured as
\begin{equation}
{\rm{RM}} = \frac{d \chi}{d \lambda^2}.
\label{rm}
\end{equation}
 
\citet{burn66} was the first to describe Faraday rotation in the more complex
situation of mixed emission and rotation, and we adopt his analysis. The operation of Faraday rotation, expressed in Equation~\ref{delta_chi}, is, of course, unchanged, but Burn introduced Faraday depth, $\phi$, a quantity analogous to ${\rm{RM}}$, defined as 
\begin{equation}
{\phi(r)}={{0.812}{\thinspace}{\int^0_r}{n_e}{\thinspace}{B_{||}}{\thinspace}dl}, 
\label{fd}
\end{equation}
where the integral is now calculated only along the line of sight from an emitting volume-element at a distance, $r$, from the observer, not through the entire magneto-ionic material in that direction{\footnote{A magnetic field, $B_{||}$, directed towards the observer, is, by convention, positive. For the Faraday depth, as defined in Equation~\ref{fd}, to be positive, $r$ must be defined with its origin at the point of emission, not at the observer.}. Every emitting volume along the line of sight has associated with it a value of $\phi$, and the observed polarized signal, ${\bf{P}}{(\lambda^2)}$, at any wavelength is the integrated sum of the Faraday-rotated emission at all Faraday depths:
\begin{equation}
{\bf{P}}{(\lambda^2)}= {{\int}^{\infty}_{-\infty}}{\bf{P}}{(\phi)}
{e^{2i{\lambda}^{2}{\phi}}}{\thinspace}d{\phi}.
\end{equation}
This has the form of a Fourier transform, and \citet{burn66} defined
the Faraday dispersion function ${\bf{F}}{(\phi)}$ as the Fourier conjugate of
${\bf{P}}{(\lambda^2)}$,
\begin{equation}
{\bf{F}}{(\phi)}= {{\int}^{\infty}_{-\infty}}{\bf{P}}{(\lambda^2)}
{e^{-2i{\lambda}^{2}{\phi}}}{\thinspace}d{\lambda^2}.
\end{equation}

When \citet{burn66} laid out these relationships they could not be implemented
because radio telescope technology and computing were not adequate for the
collection and analysis of the required data. Four decades toppled those
barriers, and \citet{bren05} developed Rotation Measure (RM) Synthesis on the
basis of Burn's equations. The technique has since been applied extensively to
data from aperture-synthesis telescopes, starting with \citet{debr05}.

The Global Magneto-Ionic Medium Survey (GMIMS) has set out to provide the data
for an improved understanding of the three-dimensional magnetic field of the
Galaxy, by mapping polarized emission over the entire sky, both the Northern and
Southern hemispheres \citep{woll09}. The Galactic polarized emission fills the
sky, with structure on all scales, and the only tools able to measure this
extended structure are single-antenna radio telescopes. GMIMS is applying RM
Synthesis for the first time to data from such telescopes. The aim is full
coverage from 300 to 1800\,MHz, with many narrow frequency channels. GMIMS
aspires beyond surveys of polarized emission, to produce surveys of Faraday
depth. When complete, the GMIMS dataset will provide a resolution in angle of
order $1^{\circ}$, and, after RM Synthesis, a resolution in Faraday depth of
order 5\,${\rm{rad}}\thinspace{\rm{m}}^{-2}$ with a sensitivity to structures in
Faraday depth space as large as 110\,${\rm{rad}}\thinspace{\rm{m}}^{-2}$. For
technical reasons the frequency band has been divided into three sub-bands,
300--800, 800--1300, and 1300--1800\,MHz (where the frequency boundaries are
approximate). The sky naturally divides into North and South, so the entire
project will comprise six {\it{component}} surveys. Observations for two
component surveys in the South (300--870\,MHz and 1300--1800\,MHz) have been
completed with the Parkes 64-m Telescope; data for 300--480\,MHz, over the
declination range  ${-90^{\circ}}\leq{\delta}\leq{20^{\circ}}$, have been
published \citep{woll19} and are now publicly available{\footnote{We denote
these two surveys by the following names and abbreviations: GMIMS Low-Band South
(GMIMS-LBS) and GMIMS High-Band South (GMIMS-HBS). The present survey is GMIMS
High-Band North (GMIMS-HBN). No mid-band surveys have been completed yet.}. 

Here we describe a GMIMS component survey of the Northern sky, covering 1280 to
1750\,MHz, and spanning declinations
${-30^{\circ}}\leq{\delta}\leq{+87^{\circ}}$, observed using the John A. Galt
Telescope (diameter 25.6\,m) at the Dominion Radio Astrophysical Observatory
(DRAO). We present the survey data, which are now being made available to the
astronomical community. The data described here have already been used to study
the two brightest polarized regions of the Northern sky, the North Polar Spur
\citep{sun15}, and the Fan Region \citep{hill17}. A region of complex polarized
emission was analyzed by \citet{woll10b}. \citet{dick19} applied moment
techniques to the data described here, and to the GMIMS-LBS data presented by
\citet{woll19}.

In Section~\ref{telescope} we describe the telescope, the receiver, and the observations. Section~\ref{datared} provides a detailed description of data processing. In Section~\ref{quality} we examine the quality of the data from the survey by comparison with existing data. Section~\ref{results} presents the results and a selection of the data and describes a few scientific outcomes and possibilities.

\section{Telescope, Receiver, and Observations}
\label{telescope}

We list observational details of the survey of polarized emission in
Table~\ref{specs}. The characteristics of the Faraday depth cube, the principal
output from this work, are given in Table~\ref{fd_specs} in
Section~\ref{results}.

\begin{table}[!h]
\caption[]{Parameters of the  polarization survey}
\label{specs}
\begin{center}
\begin{tabular}{ll}
\hline
Antenna diameter             & 25.6~m \\
Feed                         & dual circular polarization \\
Frequency coverage (observed) & 1277 to 1762\,MHz \\
Frequency coverage (usable data) & 1280 to 1750\,MHz \\
System temperature           & 140\,K \\
Angular resolution           & 38.5 to 28.1 arcmin \\
Frequency resolution         & 485~MHz/2048 = 236.8\,kHz \\
Coverage (declination)  & ${-30^{\circ}} < {\delta} < {+87^{\circ}}$ (J2000) \\
Coverage (right ascension) & ${0^{h} < RA} < {24^{h}}$ (J2000) \\
Completeness of spatial sampling & 95\% of Full Nyquist \\
Observation dates            & 2008 April to 2012 February \\
Data loss to RFI{\thinspace}$^{\dagger}$ & $\sim{30}$\% \\
Intensity calibration        & absolute \\
Angle reference              & Fan Region (see text) \\
\end{tabular}
\end{center}
~~~~~~~~$^{\dagger}$ RFI = radio frequency interference
\end{table}

\subsection{Telescope and Receiver}
\label{tel}

The receiver and polarimeter have been described in detail by
\citet{woll10a} and only an outline is given here.

The Galt Telescope is a paraboloidal reflector, of diameter
25.6{\thinspace}m. It was equipped with a feed and receiver accepting both hands
of circular polarization in a passband 1277 to 1762~MHz (the
final bandwidth of the published data is slightly smaller in extent - see
Table~\ref{specs}). A noise signal was coupled equally into both receiver
channels with a duty cycle of 50\%; its intensity was $\sim$46{\thinspace}K; 
system temperature, including the contribution from the calibration noise
signal, was $\sim$140{\thinspace}K. The calibration noise source was switched at
a 25\,Hz rate, and the polarimeter measured all inputs relative to the
calibration signal. Observations of calibration sources were made relative
to the injected noise signal, as were the scan observations that make up the
survey - see Section~\ref{cal} for details of this process, which is central to
the survey technique.

The polarimeter used commercial Field Programmable Gate Array (FPGA) circuit
modules equipped with 8-bit analog-to-digital converters. The two inputs were
digitized and processed with a Fast Fourier Transform routine to produce two spectra.  From the
left- and right-hand circular polarization inputs, L and R, four data products
LL*, RR*, LR*, and RL*, were formed (* denotes the complex conjugate). The FPGA polarimeter had a maximum clock
rate of 1 GHz, but the digitizer was clocked at 970 MHz to give an overall
bandwidth of 485~MHz with 2048 output channels of width 236.8~kHz.

The well-known advantage of using circularly polarized receivers to measure
linear polarization is that $Q$ and $U$ can be measured using cross correlation
\citep{mcco06,robi18}. In this implementation of polarimetry, Stokes vector
${I}={0.5({\rm{LL^{*} + RR^{*}}})}$, and Stokes $Q$ and $U$ equate to LR* and
RL* respectively. 

\subsection{Observations}
\label{obs}

The observations were made between 2008 April and 2012 February. The entire
survey was observed with the telescope moving up and down the meridian at 52.5
arcmin/min. This motion, together with rotation of the Earth, produced diagonal
tracks across the equatorial coordinate grid. We use the term {\it{scan}} to
denote the observation along one such track, and the scan is our basic unit of
data; we never deal with smaller units of data. Up and down scans slowly
produced a set of interlaced observations across the sky.  

Half the scans ran between declination $-30^{\circ}$ and $+87^{\circ}$ with
alternating scans running between $-30^{\circ}$ and $+60^{\circ}$ to avoid
oversampling near the North Celestial Pole (NCP). The NCP itself is not
accessible with this equatorially mounted telescope, imposing a Northern limit
of declination $+87^{\circ}$ on the survey. The southern limit, declination
$-30^{\circ}$, was set by the latitude of the Observatory and the elevation
limit of the telescope. A list of scans, the {\it{scan library}}, with
pre-determined starting right ascensions was established. Scans were set $12'$
apart in right ascension to ensure full sampling (with a beamwidth of $30'$ to
$40'$). A programming error led to a spacing of $24'$ in part of the survey, but
this error did not seriously affect sampling. Scans were chosen from the library
in a random sequence as part of a strategy to minimize systematic effects. Scans
were made only at night to avoid effects from solar emission received in the far
sidelobes, where the instrumental polarization can be as high as 50\%,
converting the unpolarized emission from the Sun into apparently polarized
emission. Spurious polarized emission from the Sun can dominate the Milky Way
polarized signal at these frequencies.  

To calibrate the intensity scale of the survey, observations of one of four
strong sources (Cassiopeia A, Cygnus A, Virgo A, and Taurus A) were made before
and after each night-time observation. The calibration sources are essentially
point sources at the angular resolution of the telescope. Although polarized
when seen at high resolution, these sources are effectively unpolarized to high
accuracy when observed with our beam. Their high intensity dominated any
sidelobe pickup, so these observations could be made in the daytime. 

\subsection{Raw Data}
\label{raw}

At the end of the observing time available for this survey in 2012
February, a total of 3536 individual sky scans had been recorded, just short of
the goal of 3600 scans.  The missing scans were not confined to any specific
part of the sky, so, with $12'$ spacing between scans and a beamwidth of $30'$
to $40'$, the sky coverage approached full Nyquist sampling.  However, as
explained below, some scans were rejected in later processing stages, which did
affect the overall sky coverage to a small extent.

\section{Data Reduction}
\label{datared}

Figure~\ref{pipeline} provides a schematic diagram of the data
reduction pipeline.  This pipeline is described to some extent in \citet{woll10a}.  Here we outline in detail the steps taken to convert the raw
scans into data cubes suitable for scientific analysis.  Each scan
carried four correlation products, RR*, LL*, LR* and RL*, and these data
were carried through the pipeline independently.

\begin{figure}[h!]
\centering
\includegraphics[bb = 180 0 750 500,width=0.9\columnwidth,clip=]{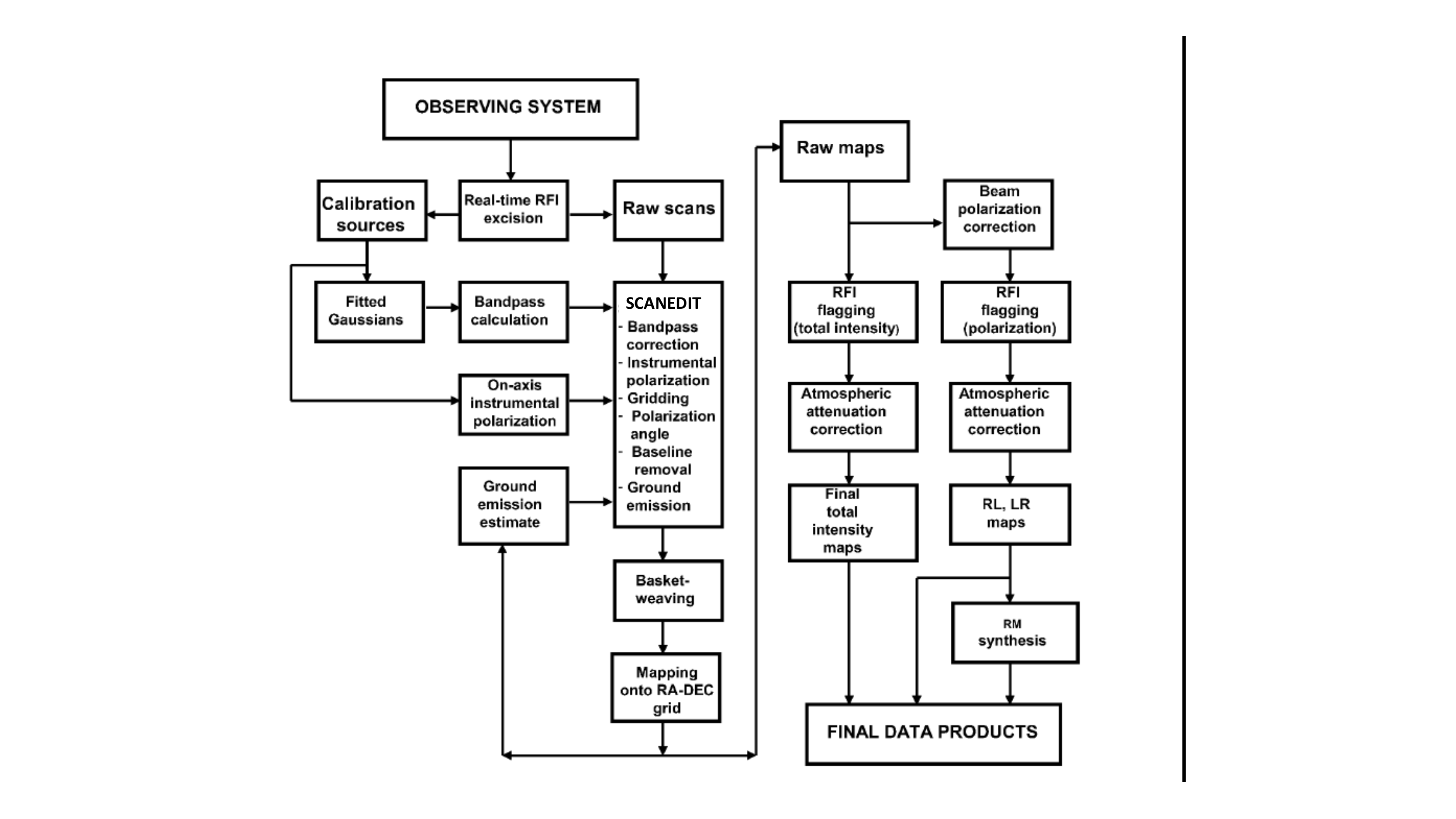}
 \caption{Schematic of the data reduction pipeline. SCANEDIT is a processing routine written for this work.}
 \label{pipeline}
\end{figure}

\subsection{Radio Frequency Interference}
\label{rfi}

Most of the observing frequencies for our survey lie outside the bands protected
for radio astronomy. The DRAO site is well protected against Radio Frequency
Interference (RFI) of terrestrial origin; it is protected physically by
surrounding mountains and administratively by various levels of government. RFI
from satellites is untouched by these measures, and remained a serious problem
with our observations. Two stages of RFI mitigation were included in the
real-time data-acquisition process: the first flagged strong, time-variable
signals, and the second employed a median filter in the frequency domain which
discarded data points lying outside a pre-determined window around the median.
Further RFI flagging was done in the final stages of the data processing
pipeline (see Section~\ref{final} and~\ref{rmsynth}). Overall data loss to RFI
was of the order of 30\%.

\subsection{Calibration}
\label{cal}

Each calibration observation consisted of a raster map of an area
$2^{\circ}{\times}2^{\circ}$ centered on the calibrator. A 2-D Gaussian above a
twisted-plane background was fitted to the observation at each frequency to
provide an amplitude, and the derived amplitude was corrected for atmospheric
attenuation using the equations of \citet{gibb86}. Prior to calibration the data
were in units of the calibration signal.  The calibration sources, with known
flux densities and spectra, provided the information to convert the data units
to Janskys.  Values of flux density, $S$, and spectral index, $\alpha$, (where
${S}\propto{\nu^{\alpha}}$), were taken from the VLSS Bright Source Spectral
Calculator
\citep{helm08}{\footnote{https://lda10g.alliance.unm.edu/calspec/calspec.html}}.
These flux densities are on the scale established by \citet{baar77}, but extend
that work with data at lower frequencies. Table~\ref{calibrators} gives these
``literature'' values, together with our adopted self-consistent values for
these parameters, based on a set of observations made before our survey
observations began. Both Cas\,A and Tau\,A are known to be declining in flux
density, Cas\,A at 0.6 to 0.7\% per year \citep{reic00} and Tau\,A at 0.167\%
per year {\citep{alle85}. The value for Tau\,A from the VLSS Bright Source
Spectral Calculator is consistent with the \citet{baar77} value allowing for an
annual decline of 0.167\% over 30 years. The somewhat lower value from our
measurements may indicate a faster decline, but that question is beyond the
scope of this paper.

\begin{table*}
\caption{Primary and Secondary Gain Calibrators}
\label{calibrators}
\begin{center}
\leavevmode
\begin{tabular}{lccccc} 
\hline
Name & \multispan{2} Literature\,value & \multispan{2} Adopted\,value & Notes \\      & Flux density & spectral index & Flux density & spectral index &  \\
     & (Jy) at 1.4\,GHz & $\alpha$ & (Jy) at 1.4\,GHz & $\alpha$ & \\
\hline
Cyg A & $1579$ & $-1.02$ & 1589 & $-$1.07 & 1,3 \\
Tau A & $908$  & $-0.29$ & 848  & $-$0.27 & 1,3 \\
Vir A & $208$  & $-0.83$ & 207  & $-$0.90 & 1,3 \\
Cas A & $2442$ & $-0.78$ & 1861 & $-$0.77 & 2,4 \\
\hline
\hline
\end{tabular}
\end{center}
Notes: 1 - primary calibrator, 2 - secondary calibrator, 3 - flux density and
spectral index taken from VLSS Bright Source Spectral Calibrator \citep{helm08}, 4 - flux density of Cas\,A decreases with time -
literature flux density is for 1980, adopted flux density for epoch 2008-2012.
\hrule
\end{table*}

The calibrations provided corrections for the instrumental bandpass, and allowed
correction of the small gain difference between LL* and RR* channels. The
bandpass was very stable through the course of the survey, and there was no
significant variation of the results obtained from different calibrators.

Each night's observations were preceded by an observation of one of the four
calibration sources, and followed by a similar observation of another. All data
were recorded in units of the injected noise signal, which was running
continuously, but, after applying the calibration, the scans were in units of
Janskys, and the intensity of the injected noise signal became irrelevant. We
did not rely on long-term stability of the noise diode; all that was required of
it was that it be stable over the course of one night's observations with their
attached calibrations. In fact the noise diode output did vary slowly over the
three years of the survey (by +13\% and $-6\%$), but this variation was so slow
that it did not contribute significant error.

Since the calibration sources were unpolarized (see Section~\ref{obs}), the
calibrations could also be used to correct for on-axis polarization leakage.
This instrumental effect arose from signal leakage between L and R, occurring in
the feed and attached waveguide devices, extremely stable metal structures. No
changes in this leakage were expected, or detected over the 3.8-year period of
observations. (Note that this step corrected for ``leakage'' between R and L
channels in the feed and polarization transducer, but did not correct for
instrumental polarization across the telescope beam. Correction for the latter
effect was made later - see Section~\ref{instpol}). While there might have been
some spurious polarized signal from the Sun in the sidelobes during the daytime
calibration observations, the calibration sources are very strong and their
emission dominated sidelobe effects; the baseline removal incorporated into the
fitting routine further diminished any sidelobe contributions. 

Polarization angle was calibrated with observations of 3C286 and
3C270 in October 2007 (see Figure~6 of \citealp{woll10a}). Further observations
of 3C286 in November 2012 revealed no significant change. However, this calibration was later revised using a new calibration technique we developed that has more general application (for a full explanation see Section~\ref{angle_tests}).

\subsection{Processing Individual Scans}
\label{scanedit}

We developed an interactive tool, SCANEDIT, for processing individual scans.
Every one of some 3500 scans was inspected for data quality as the
data came off the telescope. This program was the  principal tool for detecting
receiver and polarimeter malfunction. At this stage some scans were discarded
and observed again. In the later data processing phase,  bandpass and
instrumental polarization corrections, derived from the calibration
observations, were applied. A ground emission and atmospheric emission profile
deduced from preliminary maps was also subtracted (see Section~\ref{ground}).

\subsection{Basketweaving}
\label{basket}

The individually calibrated scans were combined to produce all-sky
RR*/LL*/LR*/RL* data cubes.  A key step in that process was ``basketweaving'',
where all scans were interleaved, and crossing points between individual scans
used to find offset values for each scan, so that small systematic variations
between scans could be minimized. This was an iterative process, applied
to the entire survey region on a channel-by-channel basis. Of the data
processing steps, basketweaving placed the heaviest demands on computing
resources {\footnote{Basketweaving used the supercomputing resources provided by
{\it WestGrid}, which is one node of {\it Compute Canada}'s High Performance
Computing facilities, and by the Centre for High Performance Computing in
Cape Town, South Africa.}}. 

Data products RR*, LL*, RL*, and LR* were processed separately by the
basketweaving algorithm, closely following the procedure of \citet{hasl74}. The
algorithm compared the signal levels along each scan with the signal levels of
all other scans that crossed this scan. Each scan will have some variations in
the baseline that are systematic on time scales of minutes or hours, but these
variations become random on time scales of weeks or months. When the baseline of
a single scan was compared to the baselines of hundreds of other scans, it could
be assumed that the baseline variations of all the other scans averaged to zero.

For each scan, the differences between data points along this scan and all the
overlapping data points from the crossed scans were calculated. Data points had
to be within $48'$ of each other to be considered overlapping. There
were usually data points approximately every $12'$ in declination along a scan,
but, to remove noise and deal with outliers, these differences were binned with
a bin width of $5^{\circ}$ in declination for six iterations, then 
$2.5^{\circ}$ for a further four iterations. We used spline interpolation
between bins. When calculating the differences, the basketweaving algorithm
discarded the lowest and highest 1\% of all differences for each overlap region.
This effectively prevented RFI from affecting baselines.

At the end of the basketweaving process the determined offsets were subtracted
from each scan. The offsets in LR* and RL* were determined over very large areas
of sky, of the order of $10^4$ square degrees. We can expect $Q$ and $U$ to
average to zero over such large areas, so no sky signal was lost. That statement
is acceptably correct in our frequency range. However, at higher frequencies
where Faraday rotation is negligible, it may no longer be true. For example, the
23\,GHz data of \citet{benn13} show polarization angle changing slowly and
smoothly with sky position.

For total-intensity data (RR* and LL*), offset removal as the last stage of
basketweaving had a more serious effect: the sky minimum at each frequency was
subtracted. The incorrect zero level means that the total-intensity data cannot be used directly for computing fractional polarization or spectral indices.

\subsection{Gridding}

Maps were made from the data after basketweaving in order to assess data
quality. Scan values falling within a square of size $12'$ on an equatorial grid
were averaged, and linear interpolation filled missing values. The products at
this stage of the pipeline were considered ``Raw Maps'', and this point is so
marked in Figure~\ref{pipeline}. The product from this stage was a set of data
cubes of RR*, LL*, RL*, and LR*.

\subsection{Ground Radiation}
\label{ground}

In the polarization channels LR* and RL*, the signal received by the telescope
was a vector combination of polarized signal from the sky, instrumental
polarization, and polarized ground emission. Instrumental polarization was
removed by the basketweaving process, but ground emission remained in the data.
Radiation from the ground entered the feed through the spillover sidelobes, which
usually have strong spurious polarization, and ground radiation reached the
receiver as a signal that appeared to be strongly polarized. It was therefore
essential to remove the effects of ground radiation from all four polarimeter
data products, not just from the total-intensity data. At the zenith, the ground
contribution to total intensity was about 5\,K. The polarized intensity of the
ground contribution was low at the zenith, but rose with increasing zenith
angle, reaching a level of ${\sim}0.3$\,K at 1.4\,GHz. This is about half the
polarized intensity of the brightest polarized features in the sky, and it
obviously had to be removed. Ground radiation did not vary with time.

We proceeded on the assumption that, across a large area of sky, the sky
polarization angle will take on a wide range of values and $Q$ and $U$ will
average to nearly zero. We used the right ascension range $8.5^{\rm{h}}$ to
$12^{\rm{h}}$, where we know that the polarized emission is low \citep{woll06},
and we averaged in right ascension. The result defined the ground emission
correction  as a function of declination and frequency. This correction was
determined channel by channel, without any smoothing in frequency, and applied
in the same way.  Removal of ground radiation was an iterative process: as maps
produced from the data pipeline gradually improved, the ground emission profile
improved in accuracy.  

In total-intensity channels, LL* and RR*, we used the same range of right
ascension. At each declination, we identified the lowest value of total
intensity, and plotted these minima against declination. Given the presence of
small emission features, this was not a smooth curve. We took the lower envelope
of this curve as the best estimate of the ground contribution as a function of
declination, but acknowledge that a small amount of Galactic signal may have
remained in this estimate. The total-intensity profiles also include atmospheric
emission ($\sim$2\,K at the zenith at the frequencies in our band, varying as
the secant of zenith angle).

\subsection{Identifying and Eliminating Bad Data}

Inspection of gridded images after the basketweaving process revealed some bad
data. The problem was ultimately traced to a faulty cable causing variation of
the calibration signal in the L receiver channel. The resulting gain jumps, of
duration tens of minutes, generated prominent features in the gridded maps that
no number of basketweaving iterations could remove. These artefacts were, of
course, most prominent in the LL* maps but strongly affected LR* and RL* maps as
well. The problem was easily solved for total-intensity data: RR* data values
were unaffected, and affected LL* data values were simply replaced by RR* data
values at the same point in the sky. This affected the final noise level to some
extent, but was otherwise not a serious degradation.

This solution was obviously unsuited to polarization data, and three different
approaches were considered to repairing LR* and RL* data, (a) keep all
3536 scans, (b) reject all scans where this problem in the L-polarization
adversely affected the data quality, and (c) attempt to ``fix'' the problem
scans by interpolating data from nearby good scans. Option (a) was rejected
because of image quality. Option (b) was rejected because of data loss. This
left option (c).

The spectrum of data quality in affected L scans was, of course, a
continuum, and judgement had to be applied. A strategy was devised
whereby acceptable L data within $30'$ of affected data were used, with
appropriate distance weightings, to generate interpolated RL* and LR*
values.  If an insufficient number of good-quality neighbouring data
values was found (if the sum of the weights was below a threshold
value) then that scan could not be ``repaired''. Applying this
interpolation scheme once, we recovered 1287 of the 1590 scans we had
previously rejected. In a second step we considered the ``repaired''
scans as good scans, and recovered another 112 scans. We did not take
this interpolation process to a third step because we would then have
been taking data from beyond the $30'$ circle. In this way we passed a
total number of 3345 scans into the basketweaving process, giving us
about 95\% Nyquist sampling of the sky. Missing data points are distributed randomly across the sky.

\subsection{Correction for Instrumental Polarization Across the
Telescope Beam} 
\label{instpol}

After on-axis instrumental polarization had been corrected (Section~\ref{cal}),
there remained instrumental polarization across the telescope beam, arising from
feed properties, reflector properties, and aperture blockage \citep{ng05,du16}.
Instrumental polarization manifests itself as leakage of Stokes $I$ into $Q$ and
$U$. In a given direction, if the ratio of the telescope response to $Q$ and
$I$, ${Q_{tel}}/{I_{tel}}$, is non-zero, a spurious polarized signal will appear
in the $Q$ channel, and, equivalently, a non-zero ratio ${U_{tel}}/{I_{tel}}$
will have the same effect in $U$. Such spurious polarized signals will result
whenever there is strong total-power emission that fills the beam (but not when
point sources are observed). For an antenna with perfectly symmetrical 
structure, the off-axis polarization is symmetrical; ${Q_{tel}}/{I_{tel}}$ and
${U_{tel}}/{I_{tel}}$ take non-zero values, but average to zero
across the main beam. However, the Galt Telescope has three feed-support struts,
which have a strong impact on polarized radiation characteristics \citep{du16},
and the instrumental polarization averages to small, but still significant,
values. The effects could be seen in our data along the Galactic plane, a strong
extended source of emission: the central parts of the Galaxy appeared to be
strongly polarized. Typical values of spurious polarized intensity in that
region were $3\%$ of the total-power signal.

We used the Galactic center region, ${\ell}\le{30^{\circ}}$, ${|b|}\le{8^{\circ}}$, to evaluate
this instrumental effect. We modified the observed values of $Q$ and $U$,
$Q_{obs}$ and $U_{obs}$, to yield $Q_{mod}$ and $U_{mod}$, where  
\begin{equation}
{Q_{mod}}={{Q_{obs}}{+}\thinspace{g}\thinspace{I}}
\end{equation}
and
\begin{equation}
{U_{mod}}={{U_{obs}}{+}\thinspace{h}\thinspace{I}}.
\end{equation}
Factors $g$ and $h$ represent the instrumental polarization. They are small
numbers which can be positive or negative, and they vary with frequency. At a
given frequency, $g$ and $h$ are constant, while the Stokes parameters vary with
sky position. $g$ and $h$ were modified iteratively until the apparent polarized
intensity in the test region was minimized. Factors $g$ and $h$ were then used
to modify observed $Q$ and $U$ across the entire survey region. The effects were
minimal, except in areas of very bright total intensity. In such areas, spurious
polarization was suppressed by about a factor of 10. 

This procedure addressed instrumental polarization across the main beam, but did
nothing for sidelobe effects. Instrumental polarization produces a
characteristic butterfly pattern in $Q$ and $U$ sidelobes. This can be seen
around very strong point sources, but it has negligible effect on the low-level extended emission. We made no corrections for sidelobe effects.

\subsection{Absolute Calibration}
\label{abscal}

The survey is absolutely calibrated, and the results are presented as main-beam
brightness temperatures in Kelvins. The aperture efficiency of the telescope,
and equivalently the gain, was measured using Cygnus A, assuming the flux
density and spectrum given in Table~\ref{calibrators}. The measurement was made
in 2015 October, after the completion of survey observations. Details of the
measurement are given in a separate paper \citep{du16} and only an outline is
given here. 

The temperature standards used in the calibration observation were (a) a box of
absorbing foam at ambient temperature that was placed in front of and around the
feed horn, and (b) the sky temperature with the telescope pointed at the zenith.
The accuracy of such a measurement is critically dependent on the cold
temperature; the knowledge gained in the antenna study \citep{du16} was applied
to estimate contributions from ground radiation and other inputs. The result was
corroborated by a number of separate measurements. First, the antenna
temperature generated by the noise calibration signal was measured relative to
noise signals from resistors at known temperatures, one immersed in liquid
Nitrogen, one at ambient temperature, and one at ${\sim}100^{\circ}$C. Second,
losses in the feed horn, the quarter-wave plate, and other waveguide components
were measured using a network analyzer. These losses amounted to about 0.3~dB, a
power loss of $\sim6\%$.

With the aperture efficiency established, the survey data could be converted into antenna
temperatures, $T_A$, in Kelvin.
\begin{equation}
{T_A}={\frac{S{\lambda^2}}{2k\Omega}}={\frac {S{\eta_A}{A_p}}{2k}},
\label{ta}
\end{equation}
where $\Omega$ is the total solid angle of the antenna in steradians, including sidelobes, $S$ is flux density in Janskys, $\eta_A$ is the aperture efficiency, ${A_p}$ is the physical area of the telescope aperture, and $k$ is Boltzmann's constant. However, the quantity of astrophysical interest is the main-beam brightness temperature, $T_B$. 
\begin{equation} {T_B} =
{\frac{\Omega}{{\Omega_B}}}{T_A}, 
\end{equation} 
where ${\Omega}_B$ is the solid angle contained in the main beam, and it is understood that all the quantities are functions of frequency, $\nu$. The beam efficiency is
\begin{equation} 
{{\eta}_B}={\frac{\Omega_B}{\Omega}}. 
\end{equation} 
The difficulty in applying these equations lies in defining the limits of the
main beam in the calculation of ${\Omega}_B$. Some surveys are reported in units
of Full-Beam Brightness Temperature, where the limits of the full beam are taken
at carefully chosen radial distance from the axis of the main beam (this limit
is set at $3.5^{\circ}$ in the work of \citealp{reic82} and \citealp{reic86} - the
well-known Stockert survey at 1420~MHz). An alternative definition is to
consider that the first null defines the limits of the main beam. These choices
are workable for surveys at a single frequency, but are difficult to adapt to a
wideband survey like that described here. We could think of no sensible way of
defining a ``full beam'' as a function of frequency. We tried using the first
null as the limit, but that moves around quite rapidly as the frequency varies,
and adopting that definition would have added frequency structure to the results
that could not possibly come from the Galactic radio emission. Instead we
defined the ``main beam'' solid angle as the solid angle of a Gaussian whose half-width equals the measured half-power beamwidth, $\theta(\nu)$, of the telescope{\footnote{\citet{baar07} (page 117) states that a Gaussian function is a good approximation to the beam from a tapered circular aperture down to a level of about $-$20\,db (1\%).}} at frequency $\nu$. Then
\begin{equation} 
{{\Omega}_B}={1.13{\thinspace}{\theta(\nu)}^2}.
\end{equation}  
Subsequent operations on the survey data assumed that the data had been taken with a Gaussian beam. In particular, prior to the RM Synthesis operation
(Section~\ref{rmsynth}), the data were brought to a common angular resolution,
the beamwidth at the lowest frequency; that was accomplished by convolution with
a Gaussian of appropriate full width half-maximum (FWHM).

Figure~\ref{apeff} shows aperture and beam efficiencies, $\eta_A$ and $\eta_B$, across the frequency band. Calculated values of $\eta_A$ are shown, from \citet{du16}. For application to processing our survey data we fitted second-order polynomials to these data points, as shown. In Section~\ref{internal_tt} we discuss frequencies around 1500\,MHz where we see the largest deviations of calculated values from the global fit; there we present evidence that aperture efficiency near 1550\,MHz is, in fact, lower than the fitted curve, as the calculated values indicate.

\begin{figure}
   \centerline{\includegraphics[bb = 40 100 580 500,
   width=0.9\columnwidth,clip=]
   {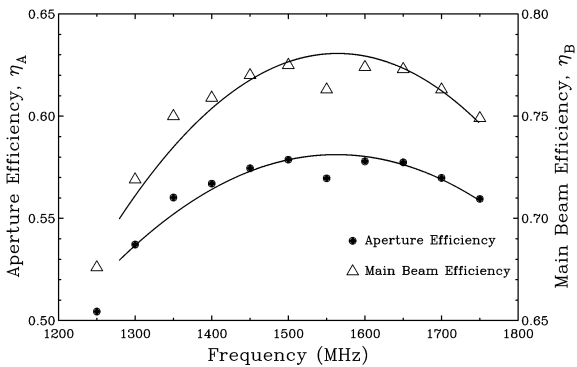}}
   \caption{Aperture and beam efficiencies, ${\eta}_A$ and ${\eta}_B$, as a 
   function of frequency. Symbols show calculated values and the curves, fitted
   to the calculated values, show the adopted function. For details see text and
   \cite{du16}. Values at 1550\,MHz, lower than the fitted curve, are addressed     in Section~\ref{internal_tt}.}
   \label{apeff}
\end{figure}

\subsection{Final Steps}
\label{final}

After completion of the basketweaving process it was clear that a few problems
remained. First, there were several obvious artefacts that were related to
declination in the total intensity maps. Second, there were distinct traces of
residual RFI in the images.

The declination-related artefacts in the total-intensity maps were
more-or-less frequency independent.  There was a stripe about
$10^{\circ}$ wide near declination $+60^{\circ}$, and a slope in level
from declination $-20^{\circ}$ to the lower limit of the survey at
$-30^{\circ}$. We assumed that these artefacts arose from an imperfect
removal of ground radiation, and we repaired them by modifying the
ground radiation function. 

The RFI remaining in the data appeared as amplitude changes along the scan
directions.  In both total-intensity and polarization data, the RFI was dealt
with in the frequency domain, but the two types of data required different
responses: in total-intensity images the excursion from apparently good data
values was always positive, while in polarization data the excursion could be
positive or negative. In polarization data, values exceeding 5 standard
deviations among data points across the frequency band were replaced by no-data
values. In the total-intensity dataset, such simple flagging removed RFI but also
flagged a large number of data points where the emission had high intensity. To
eliminate this problem a polynomial was fitted to the spectrum at each point and
subtracted from the data, effectively removing the strong emission.  Remaining
high data values were flagged, and the removed polynomial was restored. In
fitting the polynomials to the data, we ignored frequencies where RFI is always
high (for example in the GPS and other satellite bands).

As a final step the measured amplitudes of total-intensity and polarization data
were corrected for atmospheric attenuation using equations from \citet{gibb86}.

No correction was made for Faraday rotation in the ionosphere. The observations were made at night during solar minimum. At these times the ionospheric RM at DRAO is usually in the range 0.5 to 1\,${\rm{rad}}\thinspace{\rm{m}}^{-2}$, producing a rotation of only $1.6^{\circ}$ to $3.2^{\circ}$ at 1280\,MHz, and correspondingly less at higher frequencies. 

\subsection{Rotation Measure Synthesis}
\label{rmsynth}

To calculate Faraday depth (FD, $\phi$) spectra we used the 3-dimensional RM
Synthesis routines in \citet{purc20}, based on the equations in \citet{bren05},
upgraded and maintained by the Canadian Initiative for Radio Astronomy Data
Analysis (CIRADA)\footnote{RM synthesis and RM\,CLEAN code on the CIRADA github:
https://github.com/CIRADA-Tools/RM}. The code has the capability to handle
pixels flagged for RFI or lacking data, and computes a Rotation Measure Spread
Function (RMSF) unique to each pixel in the data cube that can then be used in
the RM\,CLEAN deconvolution procedure \citep{heal09}. We started with data cubes
consisting of Stokes $Q$ and $U$ channel-averaged maps, covering 1276.70 MHz to
1759.81\,MHz, smoothed to a common angular resolution of $40'$.  We averaged
five adjacent channels of the original datacube to obtain 409 channels, evenly
spaced in frequency by 1.18\,MHz. Of the 409 channels, 132 were contaminated by
RFI, including a broad frequency range spanning 1520 to 1640 MHz, and these were
not used in the RM synthesis. For the remaining 277 channels we used equal
weighting for all frequencies. 

For the frequency coverage of the survey, the resolution in Faraday depth is
approximately 150\,${\rm{rad}}\thinspace{\rm{m}}^{-2}$. This is slightly larger
in regions with missing data in the high and low frequency channels, with a
maximum value of 160\,${\rm{rad}}\thinspace{\rm{m}}^{-2}$. The RM Synthesis
parameters are summarized in Table~\ref{fd_specs} (in Section~\ref{results}) and
an example of the RMSF is shown in Figure~\ref{RMSF}. The highest frequency used
determines a maximum observable width of a broadened structure to be
${\sim}110\,{\rm{rad}}\thinspace{\rm{m}}^{-2}$, which is smaller than the width of the RMSF. The increments in $\lambda^2$
across the full frequency range are between $4.0\times10^{-5}$\,m$^2$ (high
frequencies) and $1.0\times10^{-4}$\,m$^2$ (low frequencies), corresponding to a
maximum detectable Faraday depth between
$\sim1.9\times10^{4}$\,${\rm{rad}}\thinspace{\rm{m}}^{-2}$ (low frequencies) and
$\sim4.7\times10^{4}$\,${\rm{rad}}\thinspace{\rm{m}}^{-2}$ (high frequencies).

Faraday depth spectra were calculated over the range
${-2500}\leq{\phi}\leq{2500}\,{\rm{rad}}\thinspace{\rm{m}}^{-2}$ in increments
of 5\,${\rm{rad}}\thinspace{\rm{m}}^{-2}$. This range of $\phi$ is well within
the maximum range determined by the survey parameters, and the step size
corresponds to approximately 30 samples across the FWHM of the RMSF, allowing
for smoothly displayed spectra in which features such as multiple peaks are
easily discernible. Figure~\ref{sample_spectra} shows examples of dirty and
clean spectra, together with clean components, extending over
${-1000}\leq{\phi}\leq{1000}\,{\rm{rad}}\thinspace{\rm{m}}^{-2}$. 

\begin{figure}
   \centerline{\includegraphics[bb = 1 10 1300 300,width=8.3cm,clip]
   {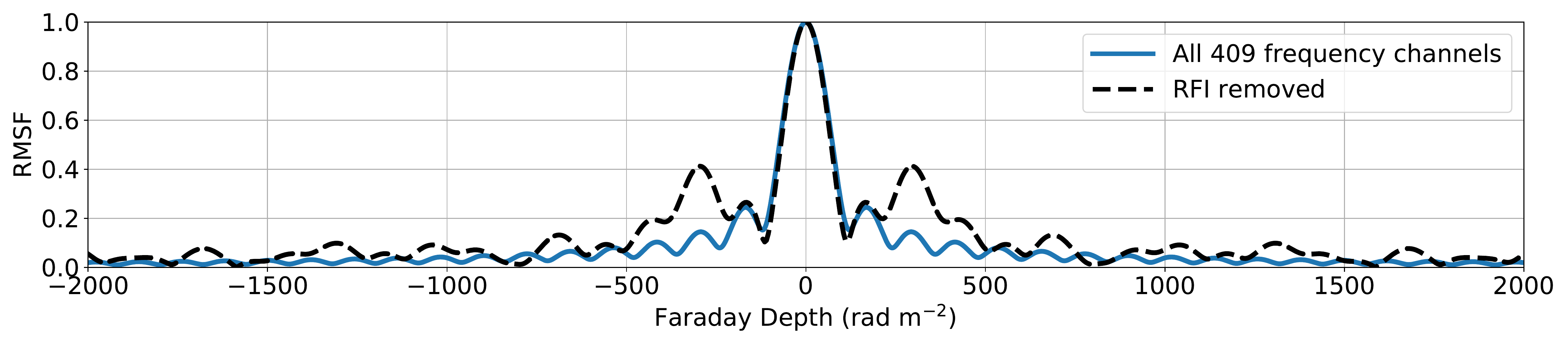}}
   \caption{A representative Rotation Measure Spread Function, before removal of RFI-affected channels from the data, and after removal of those channels.}
   \label{RMSF}
\end{figure}

\begin{figure}
   \centerline{\includegraphics[bb = 1 1 1100 400,width=8cm,clip]
   {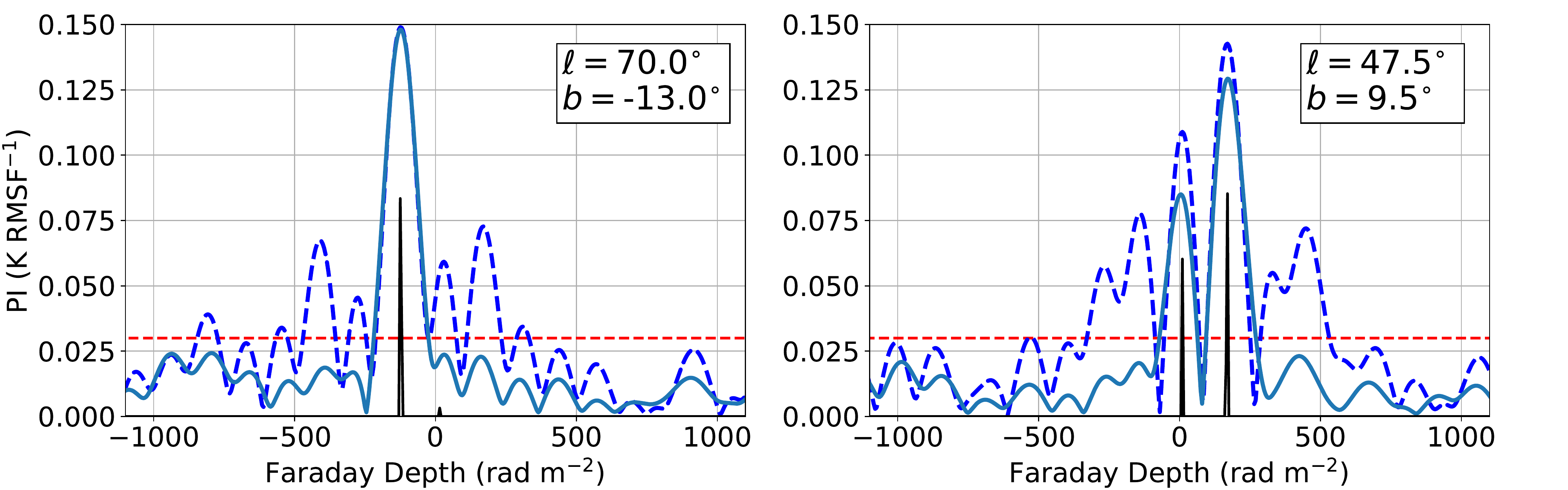}}
   \caption{Two Faraday spectra. In each plot the red dotted line indicates the CLEAN limit, 0.03\,K\,RMSF$^{-1}$. The dashed blue line shows the dirty spectrum and the solid blue line the clean spectrum. Black lines represent the CLEAN components.}
   \label{sample_spectra}
\end{figure}

A universal RM\,CLEAN threshold was determined for the entire dataset by taking
the average of the polarized intensity in the spectra beyond
$\pm$500\,${\rm{rad}}\thinspace{\rm{m}}^{-2}$, which approximates the noise
level. Taking a minimum of 5$\sigma$ for detecting a true feature yields a CLEAN
threshold of 0.03\,K\,RMSF$^{-1}$. Using an iteration increment of 10\%, the
dirty spectra were deconvolved with the RMSF provided by the RM Synthesis
procedure, down to this threshold. Many of the initial `dirty' spectra have
significant sidelobes around the main peak(s) that do not correspond to true
features. After applying RM\,CLEAN, the sidelobes are reduced to below the
threshold level. Over most of the sky there is only one peak in $\phi$, but a small fraction of
spectra show multiple peaks or broadened structures (such as in the right-hand
plot of Figure~\ref{sample_spectra}). Faraday cube characteristics are listed in
Table~\ref{fd_specs} in Section~\ref{results}. Samples from the cube, chosen to
illustrate the diversity of spectra, are shown in Figure~\ref{fd_spectra}.

\begin{figure}
   \centerline{\includegraphics[bb = 40 60 500 670,width=8cm,clip]
   {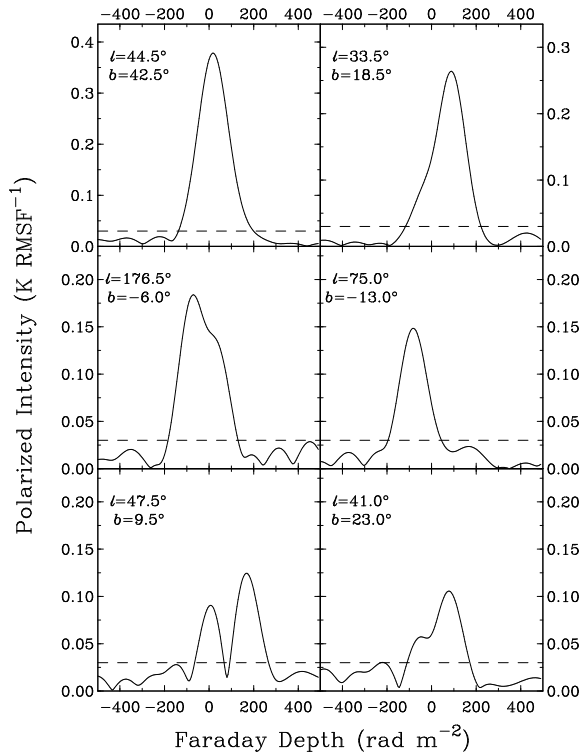}}
   \caption{Faraday depth spectra after application of RM\,CLEAN. Sky position in Galactic coordinates is shown for each spectrum. In each plot the dashed line indicates the RM\,CLEAN limit, 0.03\,K\,RMSF$^{-1}$. Different intensity scales are used for some spectra.}
   \label{fd_spectra}
\end{figure}

We inspected all spectra in the CLEANed Faraday depth cube within
${\pm}1000$\,${\rm{rad}}\thinspace{\rm{m}}^{-2}$. Features were found in some
spectra at approximately $+800$\,${\rm{rad}}\thinspace{\rm{m}}^{-2}$, and a
smaller number at $-800$\,${\rm{rad}}\thinspace{\rm{m}}^{-2}$. 
Figure~\ref{bad_fd} shows the location of all Faraday depth features beyond
${\pm}500$\,${\rm{rad}}\thinspace{\rm{m}}^{-2}$. Those at 
$-800$\,${\rm{rad}}\thinspace{\rm{m}}^{-2}$ occur at points of high total
intensity. We consider all these features to be spurious on the basis of their
apparent distribution on the sky and their narrow distribution in Faraday depth:
features at ${\pm}800$\,${\rm{rad}}\thinspace{\rm{m}}^{-2}$, confined to
declinations below $-10^{\circ}$ (see Figure~\ref{bad_fd}), are unlikely to be
related to the Galaxy. These spurious features correspond to a modulation of $Q$
and $U$ with a period of about 80\,MHz. We believe that they are by-products of
the process of determining ground radiation because they are mostly absent
between right ascensions $8^h$ and $12^{h}\,30^{m}$ where ground radiation was
evaluated (see Section~\ref{ground}). Following this investigation we decided to
publish data only over the range
${-500}\leq{\phi}\leq{500}\,{\rm{rad}}\thinspace{\rm{m}}^{-2}$.

A small modulation with a period of ${\sim}19$\,MHz is evident in the $Q$ and
$U$ images. This arises from interaction of the feed with the reflector (see
\citealp{du16} for some details). The modulation is less than a few percent
where emission is strong, but becomes fractionally more significant at low
amplitudes. The Faraday depth corresponding to this period is over
3000\,${\rm{rad}}\thinspace{\rm{m}}^{-2}$, beyond the limit of our calculations.

\begin{figure}
\centering
\includegraphics[bb = 250 20 500 700,angle=-90,width=0.8\columnwidth,clip=]
        {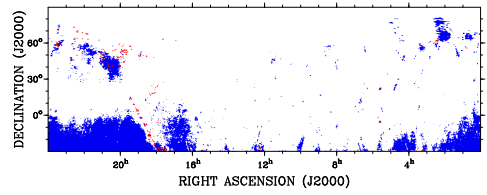}
 \caption{Locations of Faraday depth features with ${|\phi|}>{\pm}500$\,${\rm{rad}}\thinspace{\rm{m}}^{-2}$. Blue indicates features with positive $\phi$, the majority, and red indicates  features with negative $\phi$, only 1.5\% of the total. All are considered spurious.}
 \label{bad_fd}
\end{figure}

\section{Tests of Data Quality}
\label{quality}

In this section we compare our survey data to existing data, where available, and we describe some tests of internal consistency. We also estimate error in the dataset.

\subsection{The Amplitude Scale Near 1400\,MHz}
\label{ampl_tests_1400}

We wanted to compare our data against existing data, but that comparison had to
be confined to the vicinity of 1.4\,GHz, the only frequency within the range of
our survey where other datasets exist. We used the T-T plot method
\citep{cost60}, in which the intensity from one dataset was plotted against the
intensity from another dataset at the same sky position. We made T-T plots for
total intensity, $I$, and for polarized intensity. When the two datasets are at
the same frequency, the slope of the line fitted to the points gives the average
ratio between the two temperature scales.

First, we compared our $I$ data with the Stockert dataset \citep{reic82, reic86}
at 1420\,MHz (with frequency channels chosen to match the Stockert bandpass).
The comparison, over the entire range of our survey, is shown in
Figure~\ref{stockert_compare}. The fitted line shown in the figure has a slope
of 1.38. If the Stockert full-beam brightness temperatures are converted to
main-beam brightness temperatures using the full-beam and main-beam solid angles
given by \citet{reic88}, the GMIMS/Stockert ratio is 0.97. We note that the two surveys were made 30 years apart, were independently calibrated, and used different definitions of the main beam in calculating beam solid angle. We have not made any adjustments to our intensity scale. The offset of the fitted line, about $-1$\,K, arises from the basketweaving process, which has removed the sky minimum from our $I$ data. We did not attempt to correct the zero level of our $I$ data.

\begin{figure}
   \centerline{\includegraphics[bb = 50 50 600 650,width=8cm,clip]
{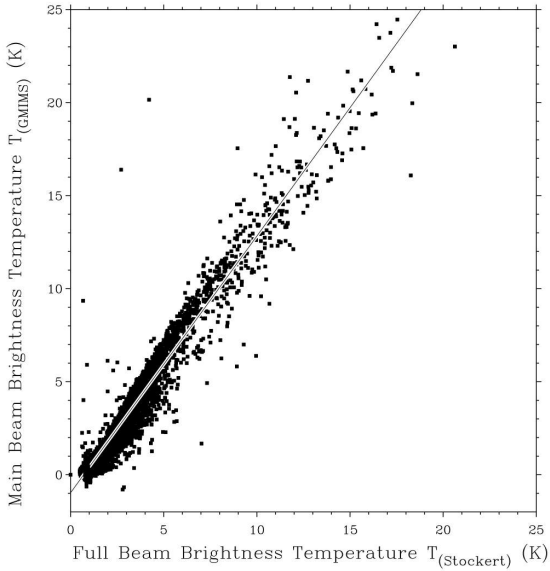}}
   \caption{Main beam brightness temperature at 1420\,MHz plotted point-by-point against full beam brightness temperature from the Stockert surveys (\citealp{reic82}, \citealp{reic86}) over the entire range of our survey. A few points at very high intensity, corresponding to small-diameter sources, are omitted. The fitted line has a slope of 1.38 and an offset of $-0.97$\,K.}
   \label{stockert_compare}
\end{figure}

The main theme of our work is a study of the polarized sky, so very relevant comparisons center on the polarized emission. We made use of the data of \citet{brou76}, a set of carefully calibrated surveys at 408, 465, 610, 820, and 1411\,MHz, made with the Dwingeloo 25-m Telescope (we refer to these five datasets collectively and separately as the {\it{Dwingeloo data}}). 
We compared our values of polarized intensity at 1411\,MHz with the 1411-MHz Dwingeloo data; the angular resolution is almost identical to ours. Our source of Dwingeloo data was a computer-readable file giving values of polarized intensity and polarization angle over most of the sky above declination $0^{\circ}$. Figure~\ref{pi_compare} shows T-T plots of polarized intensity over the two most highly polarized regions of the Northern sky, the Fan Region and the North Polar Spur. The two scales are clearly quite similar, but there is scatter in both plots and there are outliers. To quantify the comparison we have computed histograms of the ratio between the two surveys; these are shown in Figure~\ref{ratio_hist}. The histograms peak at a ratio about 0.9, implying that our polarized intensities are slightly higher than the Dwingeloo values. 
The two surveys have different sensitivity:  \citet{brou76} quote 60\,mK as the
mean error of their 1411 polarized intensities, while the noise on our data over
the equivalent band at 1411 MHz is 25\,mK. Some of the difference between the
two survey scales can be attributed to slightly different definitions of the
main beam, something we cannot make adjustments for. 

\begin{figure}
\centering
\includegraphics[bb= 40 80 570 600,angle=0, width=0.8\columnwidth,clip=]
        {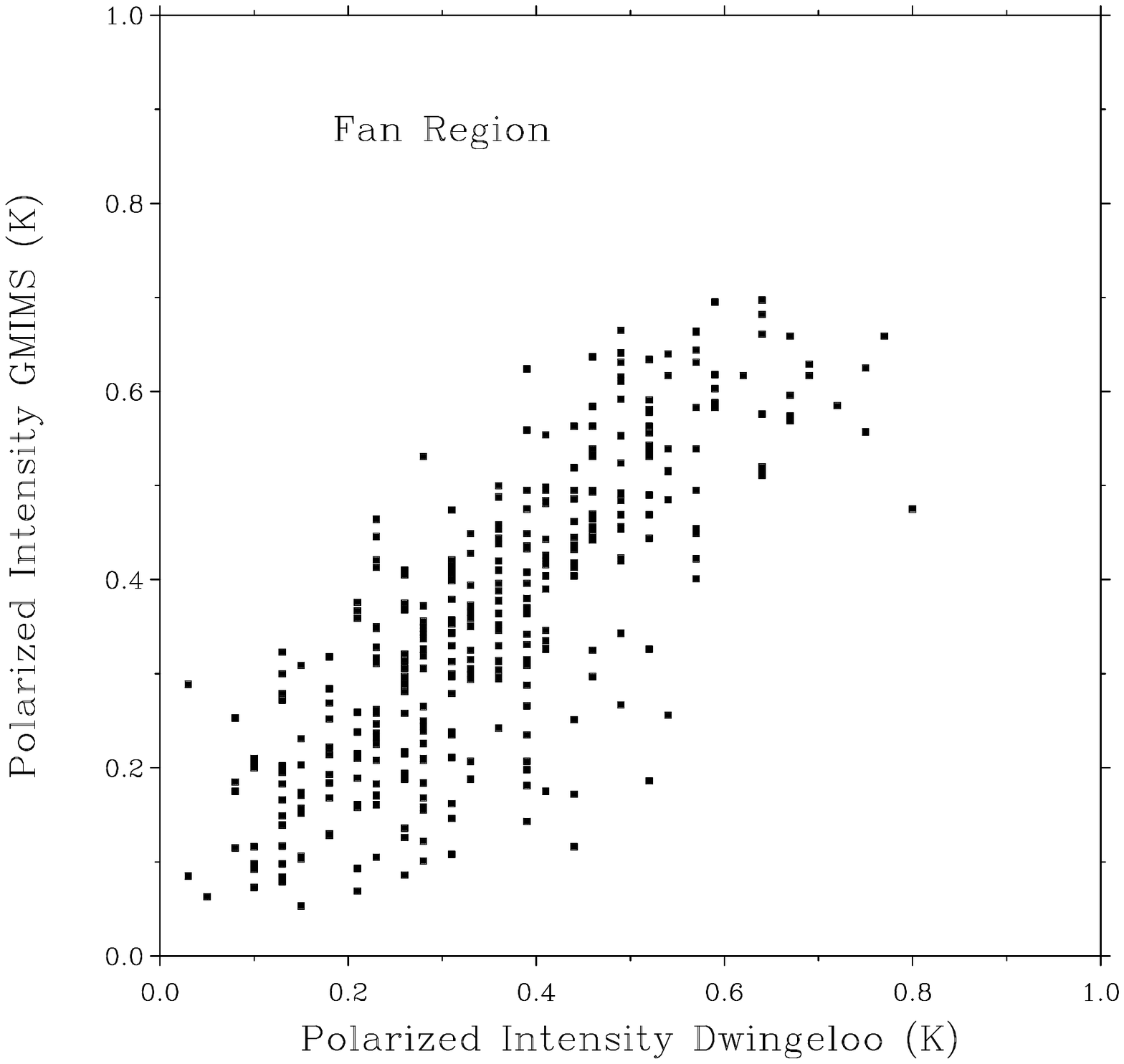}
\includegraphics[bb= 40 80 570 600,angle=0, width=0.8\columnwidth,clip=]
        {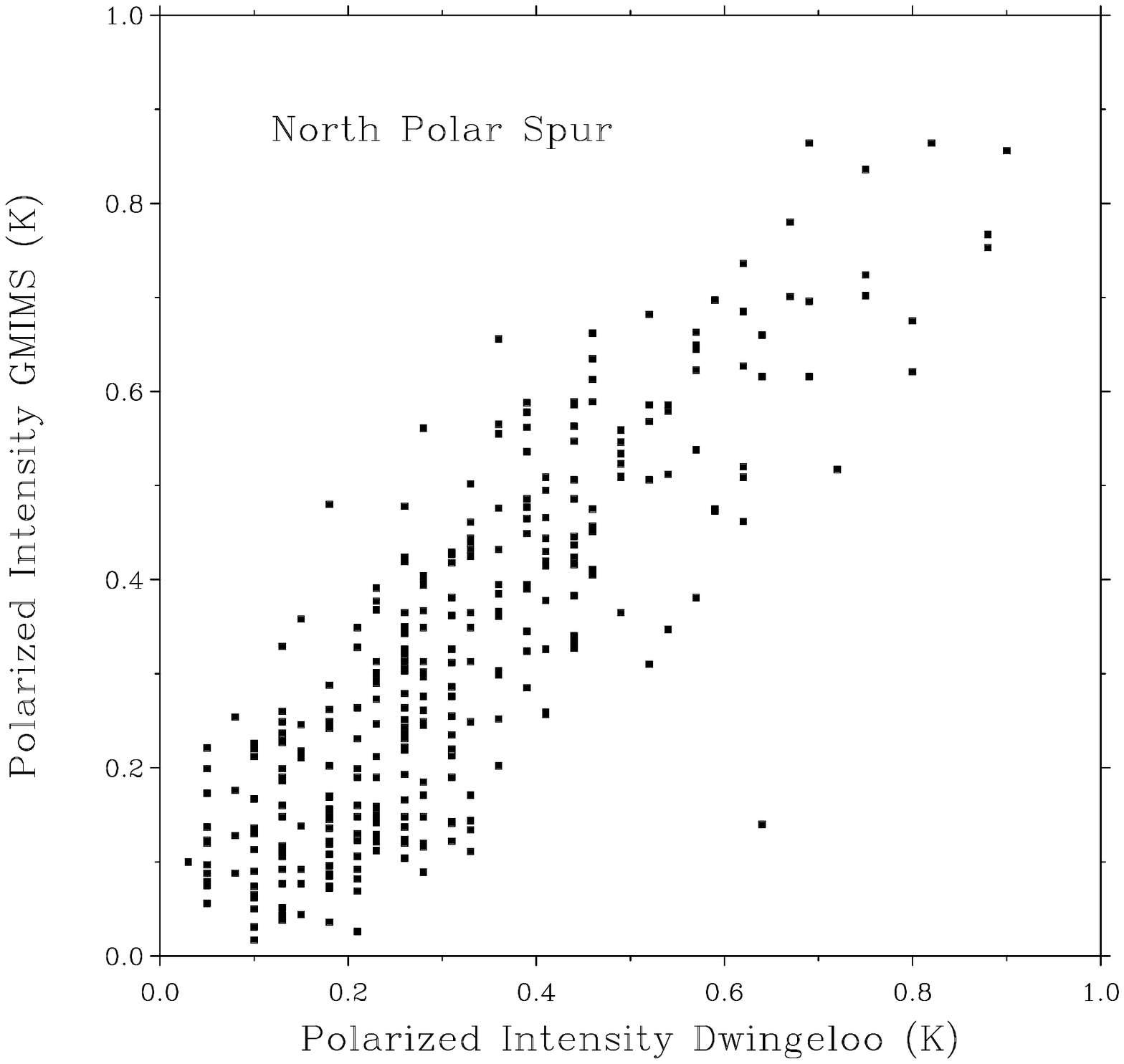}
 \caption{Polarized intensities at 1411\,MHz plotted point-by-point against the corresponding values from the Dwingeloo survey \citep{brou76}, for the Fan Region (top) and the North Polar Spur (bottom). The Fan Region data, comprising 350 sky positions, cover
${110^{\circ}}\leq{\ell}\leq{180^{\circ}},
{0^{\circ}}\leq{b}\leq{30^{\circ}}$. Data for the North Polar Spur, 319 points, cover
${-30^{\circ}}\leq{\ell}\leq{60^{\circ}},
{20^{\circ}}\leq{b}\leq{80^{\circ}}$. }
 \label{pi_compare}
\end{figure}

\begin{figure}
\centering
\includegraphics[bb = 40 80 570 650,angle=0, width=0.8\columnwidth,clip=]
        {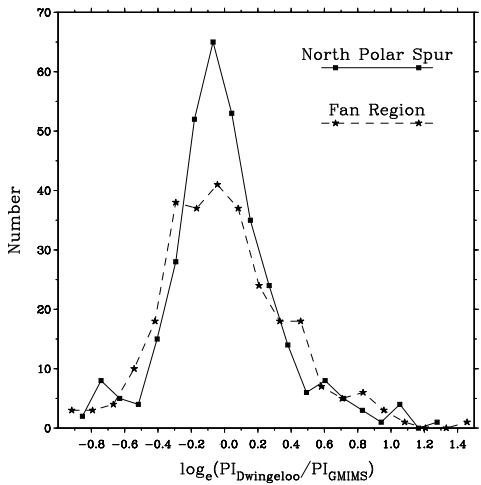}
 \caption{Histograms showing the ratio between Dwingeloo polarized intensities and GMIMS polarized intensities at 1411\,MHz. A threshold is set for both surveys at ${\rm{PI}}={0.1}\,{\rm{K}}$. The quantity plotted is 
$\rm{log_e({PI_{Dwingeloo}}/{PI_{GMIMS}})}$.}
 \label{ratio_hist}
\end{figure}

\subsection{The Amplitude Scale Across 1280 to 1750\,MHz}
\label{internal_tt}

We investigated the relative accuracy of the intensity scale of the survey using the total-intensity data. We generated $I$ maps in nine frequency bands that were relatively free of RFI - details are given in Table~\ref{bands}. Within each band, individual channels are deleted due to RFI. Consequently the bands do not appear to be evenly spaced, and appear to have different widths in frequency. There is a significant gap between 1521 and 1605\,MHz where the RFI was particularly severe.

\begin{table}[h]
\caption{Selected Bands For Internal Spectrum Consistency Investigation}
\label{bands}
\begin{center}
\leavevmode
\begin{tabular}{cccc} 
\hline
Band & Frequency range & Centre frequency & Width \\
     & (MHz)           &       (MHz)      & (MHz) \\
\hline
1    & 1289.1 - 1322.3 &      1305.7      &  33.2 \\
2    & 1322.3 - 1355.4 &      1338.9      &  33.2 \\
3    & 1355.4 - 1418.2 &      1386.8      &  62.8 \\
4    & 1418.2 - 1451.4 &      1434.8      &  33.2 \\
5    & 1451.4 - 1483.3 &      1467.3      &  32.0 \\
6    & 1483.3 - 1521.2 &      1502.3      &  37.9 \\
7    & 1605.3 - 1653.8 &      1629.6      &  48.5 \\
8    & 1653.8 - 1691.7 &      1672.8      &  37.9 \\
9    & 1691.7 - 1733.2 &      1712.4      &  41.4 \\
\hline
\hline
\end{tabular}
\end{center}
\end{table}

All-sky maps were made in the nine frequency bands. Since the sky minimum was
removed from the total-intensity data by basketweaving, we could not compute the
absolute spectral index. Instead, we computed T-T plots between pairs of
frequencies over selected areas. In this experiment, where the two input
frequencies are different, the slope of the T-T plot gives the temperature
spectral index, $\beta$, in that frequency range, where
${T_B}\propto{{\nu}^{\beta}}$. The differential T-T plot method is unaffected by zero level errors.

Our primary test was made using the Cygnus-X area. In this complex region, the
line of sight passes along the local spiral arm \citep{wend91}. There are many
\ion{H}{2} regions in Cygnus X (\citealp{knod00}; \citealp{gott12}), so there
is a large amount of thermal emission. This region shows ${\beta}\approx{-2.4}$
in the spectral index map of \citet{reic88}, computed between 408 and 1420\,MHz.
This is a lower value of $\beta$ than the surroundings, indicating a mix of
thermal and non-thermal emission. \citet{xu13} demonstrated that the thermal
emission in Cygnus X is superimposed on a spatially nearly uniform background of
non-thermal emission. We therefore expect the {\it{differential}} spectral
index, as derived from our T-T plots, to be very close to ${\beta}={-2.1}$, the
value for optically thin thermal emission.  T-T plots between the lower
frequency channels and the upper channels did indeed produce values of $\beta$
near $-2.1$, and we concluded that we could use this region as a calibrator. 
T-T plots that involved channels 4, 5, and 6, at frequencies near 1500\,MHz,
produced results that implied that intensity scales of these bands were slightly
too low. We adjusted data in channels 4, 5, and 6 upwards by factors of 1.04,
1.05 and 1.05 respectively. Figure~\ref{TT_cygx} shows T-T plots from eight
pairs of frequencies over Cygnus X, after this adjustment to the central
channels. Weighting the eight derived values of $\beta$ by the frequency
interval of each determination, we derived spectral indices for Cygnus X of
${\beta}={-2.10}$ with a scatter of ${\pm}0.08$. 

We then proceeded to derive T-T plots over an area of intense emission near the
Galactic center, just off the Galactic plane, where we expect the emission to be
predominantly non-thermal, with a steeper spectrum. The T-T plots are shown in
Figure~\ref{TT_gc} (after adjustment of central frequency channels). Averaging
using the same weighting as above, we obtained ${\beta}={{-2.50}\pm{0.09}}$.

These results are highly consistent, considering that many of these T-T plots
are made over small frequency ranges. From this experiment we conclude that the
intensity scale is well determined across the band within a few percent. If the
error in the intensity scale between 1306 and 1712\,MHz was 3\%, that would
produce a change in $\beta$ outside the range of values shown in
Figures~\ref{TT_cygx} and \ref{TT_gc}. We conclude that the probable error in
relative intensities within the band is  ${\pm}2{\%}$, after the small
adjustment of about 5\% to frequencies near band center. We note that 5\% is
within the estimated overall error in our intensity scale (see
Section\,\ref{errors}).  Examining Figure~\ref{apeff} we see that there are
departures of this order of magnitude of the calculated aperture efficiency from
the fitted second-order polynomial, especially at 1550\,MHz, and we suggest this
may be responsible for the scale discrepancy detected in the middle of the band.
T-T plots involving data at 1550\,MHz would have provided a test of this
hypothesis, but because of RFI we could not make a useful total-intensity map at that frequency.

\begin{figure}
\centering
\includegraphics[bb = 50 50 600 650,angle=0, width=1.0\columnwidth,clip=]
        {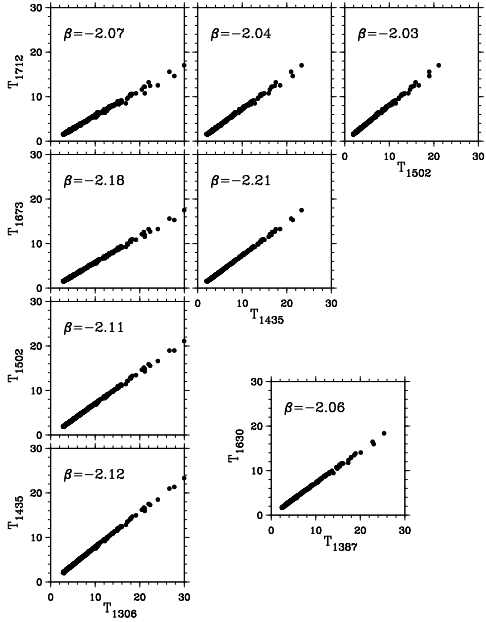}
 \caption{T-T plots made from 312 independent data points in the Cygnus-X area, covering an area of 78 square degrees defined by ${75^{\circ}}\leq{\ell}\leq{86.5^{\circ}},{-2^{\circ}}\leq{b}\leq{4^{\circ}}$.
The derived temperature spectral index, $\beta$, is shown in each plot. See text for details.}
 \label{TT_cygx}
\end{figure}

\begin{figure}
\centering
\includegraphics[bb = 50 50 600 650,angle=0, width=1.0\columnwidth,clip=]
        {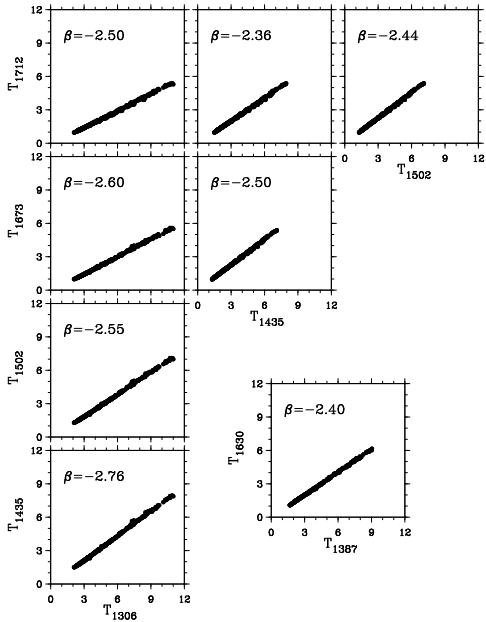}
 \caption{T-T plots made from 384 independent data points in the vicinity of the Galactic center, covering an area of 96 square degrees defined by ${2^{\circ}}\leq{\ell}\leq{13.5^{\circ}},{2.5^{\circ}}\leq{b}\leq{10^{\circ}}$.
The derived temperature spectral index, $\beta$, is shown in each plot. See text for details. }
 \label{TT_gc}
\end{figure}

\subsection{Re-evaluation of the Angle Calibration}
\label{angle_tests}

Comparison of polarization angle was possible with other data only in the
vicinity of 1400\,MHz: no surveys have been made at other frequencies in our
band. Comparison with the single-frequency survey of \citet{woll06} showed a difference of polarization angle of ${\sim}20^{\circ}$. That survey was calibrated using the 1411\,MHz Dwingeloo data \citep{brou76}; as expected, a direct comparison of the new data with the \citet{brou76} data at 1411\,MHz indicated a very similar angle offset. We had no {\it{a priori}} way of establishing which, if any, of the three surveys was correct, but in this section we develop and apply a new angle calibration technique based on the Fan Region.

The Fan Region is an area where polarization angle changes very slowly with
frequency, a fact well established from earlier polarization surveys. In
Figure~\ref{fd_compare} we show a comparison of results derived from the new
Faraday cube with a map of RM from \citet{spoe84}. 
The lower panel shows Spoelstra's result, the RM computed from the Dwingeloo data, made by fitting
observed polarization angle as a function of wavelength squared (as in
Equation~\ref{rm}) to narrow-band measurements of polarization angle at 408,
465, 610, 820, and 1411\,MHz. The upper panel shows the first moment computed
from our Faraday cube, using the equations presented by \citet{dick19}. In the
Fan region, where the Faraday depth structure is very simple, the first moment
of Faraday depth is essentially equal to the ``RM'' value that would be
calculated from our data by fitting polarization angle, $\chi$, as a function of
${\lambda}^2$. The two plots in Figure~\ref{fd_compare} are therefore quite
comparable, and they are indeed strikingly similar. The line of zero RM in the
Dwingeloo data corresponds closely to the line of zero first moment in our data,
and the two datasets are correlated: where the FD is positive the RM is
positive, and {\it{vice versa}}.  

How reliable is this comparison? Our survey is fully sampled in frequency and
angle, and the image shown in Figure~\ref{fd_compare} comprises 7200 independent
data points. In contrast, the Dwingeloo observations are sparsely sampled in
frequency and angle: the RM plot in Figure~\ref{fd_compare} is defined by
only 227 data points. Nevertheless, the polarization structure in the Fan Region
is very simple, changing quite slowly with sky position, and Faraday depths are
low, so we consider the Dwingeloo RM values in Figure~\ref{fd_compare} to be a
reliable representation of Faraday depth in the Dwingeloo frequency range.

Figure~\ref{fd_comp_plot} shows a comparison of the mean FD values (first moment of the FD spectrum) point by point with the corresponding RM values over the area ${120^{\circ}}<{\ell}<{170^{\circ}}, {0^{\circ}}<{b}<{20^{\circ}}$. The two datasets are strongly correlated, but the GMIMS FD values are larger than the Dwingeloo RM values. The plotted line in Figure~\ref{fd_comp_plot} has a slope of 2; this not a fitted line, but examination of the figure shows that it approximately represents the data. We avoid presenting a fit to these data points because we do not want to over-interpret this result (and the appearance of the plot changes slightly depending on the exact area from which data points are selected).  

\begin{figure}
\centering
\includegraphics[bb= 160 70 450 710,angle=-90, width=0.85\columnwidth,clip=]{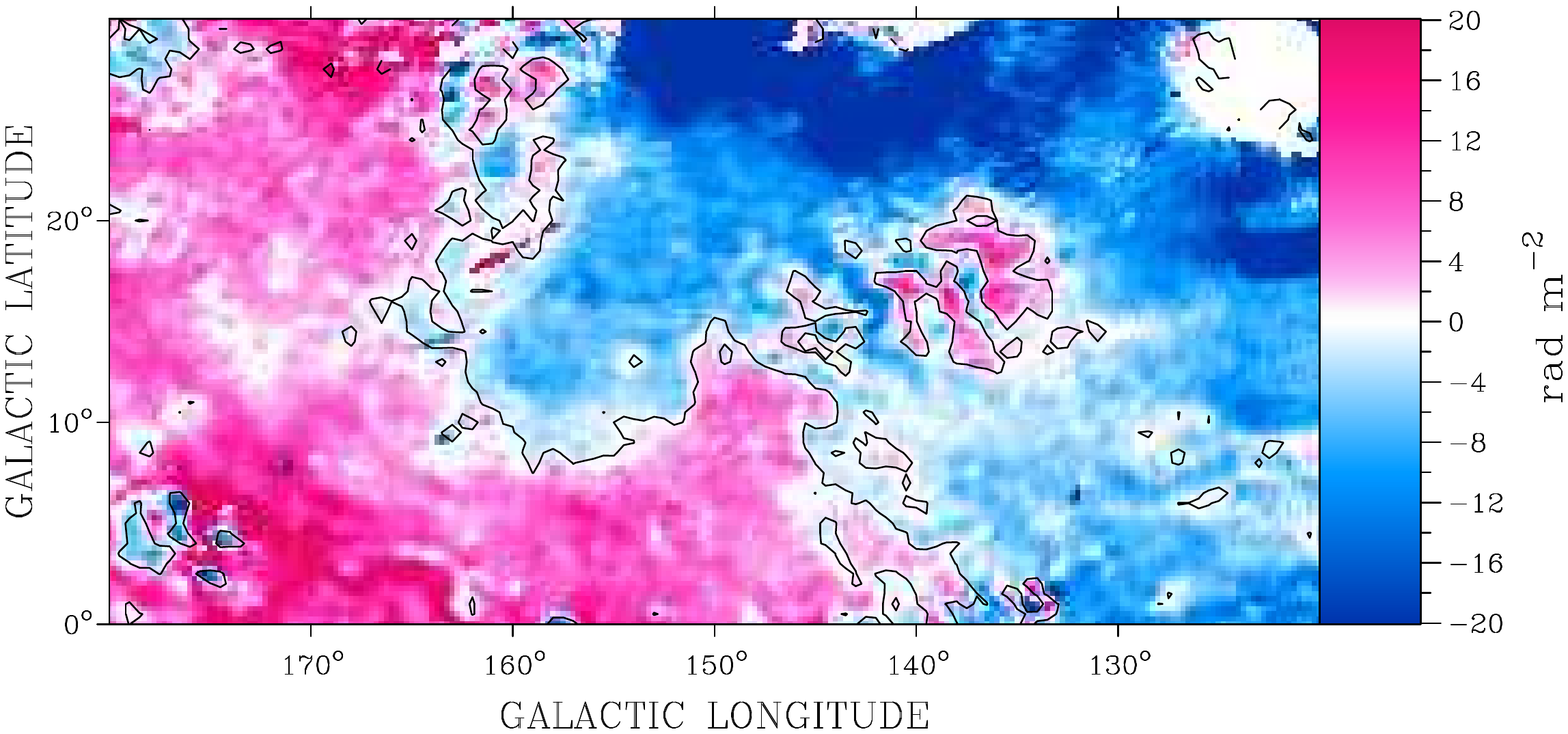}
\includegraphics[bb= 160 70 550 710,angle=-90, width=0.85\columnwidth,clip=]{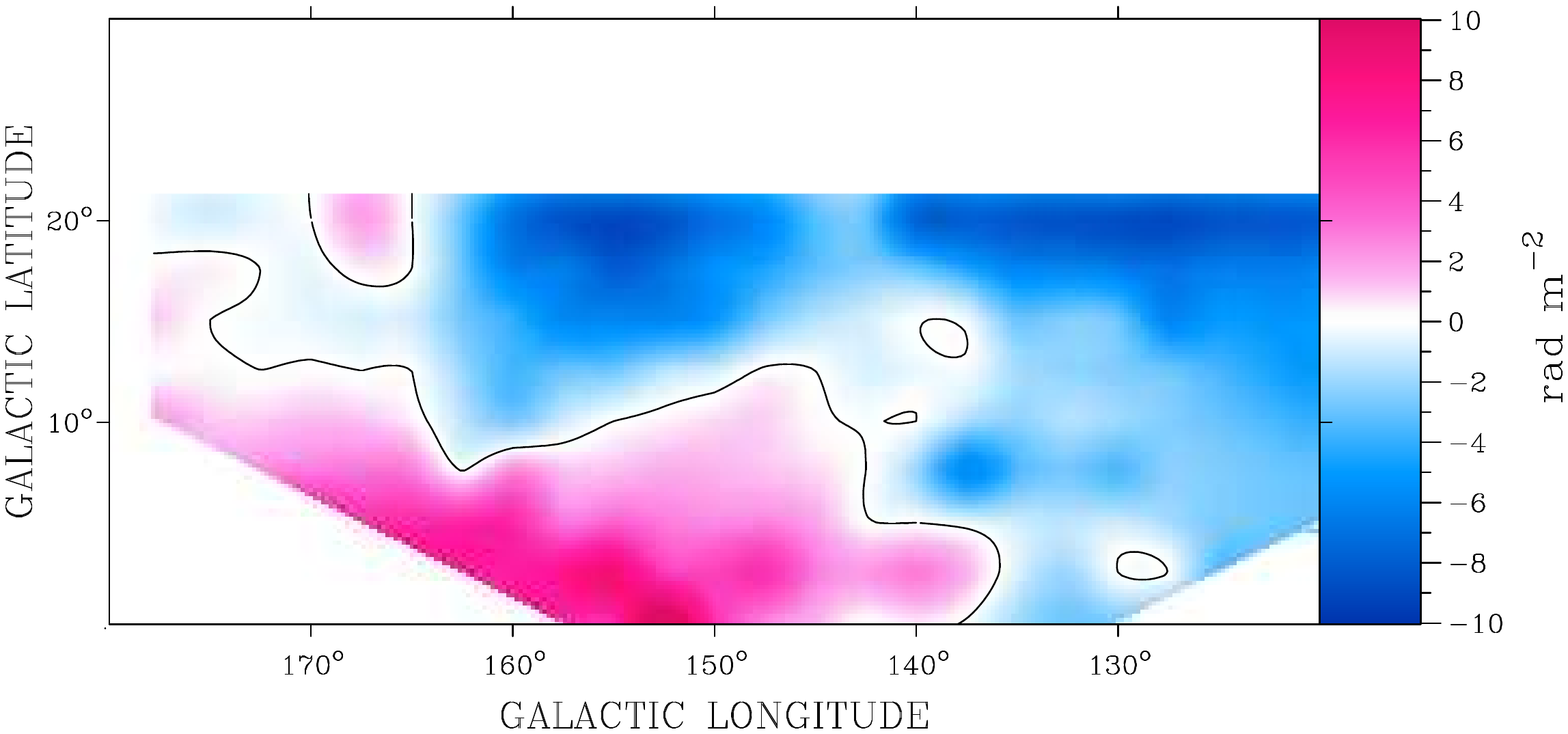}
 \caption{Top: GMIMS Moment 1 image over the Fan Region derived from the present
survey, covering 1280 to 1750\,MHz. The white region at top right is the
area north of declination $87^{\circ}$ where observations could not be made.
Bottom: Rotation Measure derived by \citet{spoe84} from data between 408 and
1411\,MHz over the same region. The bottom figure is based on data from
\citet{brou76} and is confined to the area covered by Figure\,2 of
\citet{spoe84} (calculated from the undersampled Dwingeloo data). The single
contour shown in each image corresponds to
0\,${\rm{rad}}\thinspace{\rm{m}}^{-2}$. Note that the range of the color scale
in the lower image is half that used in the upper image.}
 \label{fd_compare}
\end{figure}

\begin{figure}
   \centerline{\includegraphics[bb = 50 50 600 650,width=8cm,clip]
   {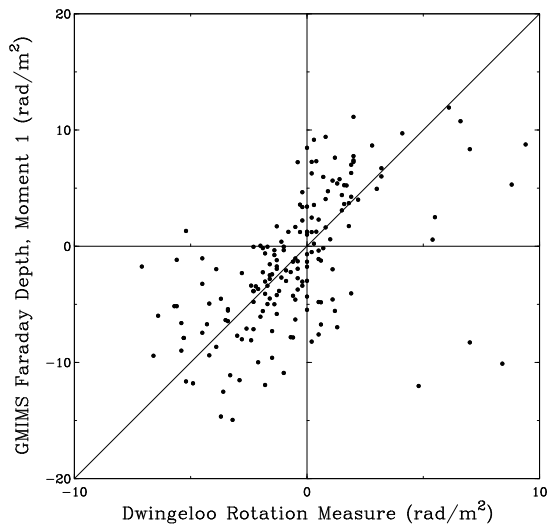}}
   \caption{GMIMS Moment 1 values plotted against Dwingeloo RM values in the Fan Region. The plot covers the area ${120^{\circ}}<{\ell}<{170^{\circ}}, {0^{\circ}}<{b}<{20^{\circ}}$. The diagonal line has slope of 2; it is not a fit to the data. There are 185 data points in this plot, the number of Dwingeloo observations. (Note that this plot covers a slightly smaller area than the comparison in Figure~\ref{fd_compare}).}
   \label{fd_comp_plot}
\end{figure}

The most straightforward interpretation of Figure~\ref{fd_comp_plot} is that the
Dwingeloo data, covering 408 to 1411\,MHz, mostly represent the nearby emission,
because polarization horizon effects at those low frequencies, where beamwidths
are large, confine the observations to the nearby magneto-ionic medium. The 
GMIMS FD data, covering the frequency range 1280 to 1750\,MHz, are sensitive
to the magneto-ionic medium over a greater range of distance. Polarized emission
from larger distances is likely to suffer more Faraday rotation than is
experienced by local emission. Detailed interpretation of this result is beyond
the scope of this paper; interpretation will be easier when polarization data
that are fully sampled in the low frequency range become available for the Fan
region from surveys presently underway with DRAO telescopes.

The comparisons that we have made suggest that there is a specific area within
the Fan Region where there is no, or very little, Faraday rotation between 408
and 1750\,MHz (and, of course, at any higher frequency). This suggests that this
area can be useful for calibration of polarization surveys in the Northern sky.
We have put this into practice to re-calibrate the polarization angle for our
survey. For this purpose we adopt the K-band (23\,GHz) data from WMAP
\citep{benn13} as the calibration standard. 

\begin{figure} 
\centering
\includegraphics[bb= 170 70 500 650,angle=-90,width=0.85\columnwidth,clip=]
{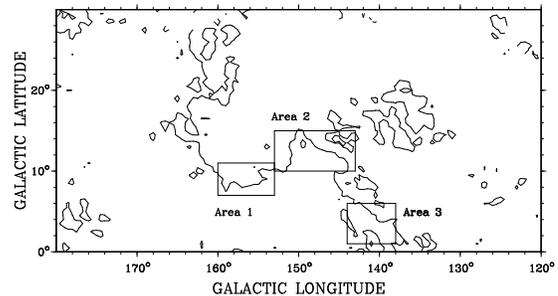} 
\caption{The area of the Fan Region, showing the contour of zero Faraday depth (from our Moment 1 data, as shown in Figure~\ref{fd_compare}). Superimposed are outlines of three regions chosen for use in calibration of  polarization angle. See text for details.}
\label{fd_zero} 
\end{figure} 

Figure~\ref{fd_zero} covers the same area as Figure~\ref{fd_compare}, and shows
the contour of zero Faraday depth. Rectangular boxes define three regions that
lie along that contour, chosen to lie below ${b}=15^{\circ}$ to capture the
highest polarized intensity from the Fan region. In
Table~\ref{fan_region_angles} we compare polarization angles in these three
regions from the present data (designated as GMIMS), the data of \citet{woll06}
(DRAO (2006)), and \citet{benn13} (WMAP). The GMIMS angles listed in
Table~\ref{fan_region_angles} were computed over the entire survey band, 1280 to
1750\,MHz, removing the range 1520 to 1605\,MHz where the data are severely
affected by RFI. For each area, and each data set, we present the average
polarization angle in that area, followed by a number in parentheses which is
the rms of the angle values in that area. In the case of GMIMS angles, the rms
is calculated over all frequencies in the survey. Taking all three regions
together, the average angle for the GMIMS data is ${-25^{\circ}}$ and the
distribution of values has an rms of ${8^{\circ}}$. We take the latter value as
our estimate of measurement error for angle in the survey, (see
Section~\ref{errors}).

\begin{table*}
\caption{Fan Region Polarization Angles}
\label{fan_region_angles}
\begin{center}
\leavevmode
\begin{tabular}{cccccccc} 
\hline
Area & Range in  & Range in & Area    & WMAP & GMIMS & DRAO (2006) & \\
     & longitude & latitude & (square~ & PA$^{\dagger}$ & PA$^{\dagger}$ & PA$^{\dagger}$ & \\
     & (degrees) & (degrees) & ~degrees) & (deg) & (deg) & (deg) & \\
\hline
1    & 153 - 160 & 7 - 11 & 28 & $-{3.5}~(5.0)^*$ & ${-24.5}~(5.9)$ & ${-3.4}~(3.5)$ & \\
2    & 143 - 153 & 10 - 15 & 50 & $-{6.2}\,(4.5)$ & ${-29.4}\,(9.2)$ & ${-2.3}\,(7.0)$ & \\
3    & 138 - 144 & 1 - 6 & 30 & ${-1.2}\,(3.4)$ & ${-19.3}\,(4.3)$ & $~~{3.1}\,(2.3)$ & \\
\hline
All  &           &       & 108 & $-{4.1}\,(4.8)$ & ${-25.2}\,(8.4)$ & ${-1.1}\,(5.8)$ & \\
\hline
\hline
\end{tabular}
\end{center}
{$^{\dagger}{\thinspace}$All polarization angles are in the astronomical reference system, with zero at the Galactic north pole increasing to the east. 

$^*{\thinspace}$Each angle value is followed by a number in parentheses, which is the rms calculated from the angle values in that area.} 

\end{table*}

The conclusion from inspection of Table~\ref{fan_region_angles} is that the
GMIMS angles differ from the WMAP angles by $21.1{\pm}{9.7}^{\circ}$, and that
our angle calibration based on 3C286 is in error by this amount. Taking this
result to one significant figure, we have added $20^{\circ}$ to our polarization
angles and re-calculated the $Q$ and $U$ data. The Faraday depth cube was
completely unaffected by this operation because the same angle offset was
applied to all frequency channels.  The systematic error of WMAP polarization
angles is $1.5^{\circ}$, plus an error up to $1^{\circ}$ dependent on polarized
intensity \citep{benn13}. In the Fan Region this error is likely to be 
${\sim}0.3^{\circ}$ (J. Weiland, private communication, 2020). These errors are
small compared to the errors in our data.

The Fan Region is unique as an angle calibrator for single-antenna polarization observations. The Fan Region has high polarized intensity, and its polarization angle is unchanging from 408\,MHz to high frequencies. No other region in the Northern sky has this combination of properties.

\subsection{Error Analysis}
\label{errors}

We discuss errors in polarized intensity and polarization angle separately.

The noise on $Q$ and $U$ images, measured in low-signal regions of images made with a channel width of 1.18\,MHz, is 45\,mK. From the known system temperature and integration time, we expect a noise level of 41\,mK rms (calculated following the method of \citealp{mcco06}). The noise on the Faraday depth cube is lower, 3.3\,mK, because the entire bandwidth participates in the determination of each data value in this cube. The theoretical estimate is 2.4\,mK, calculated on the basis of the RFI-free bandwidth that has been used in computing the Faraday depth cube.

The error in our knowledge of the flux density of Cygnus\,A is 5\% (derived from
the errors quoted by \citealp{baar77}). Beyond this is the possibilty of error
in the determination of aperture efficiency, arising in the actual measurements;
this is 3\% \citep{du16}. There is definitely additional error that must be
considered, from the application of the calibration data to individual scans and
the processes, such as basketweaving, that we have applied to the data. We
estimate this error as 5\%. Combining these errors, the probable error in
polarized intensities is 8\%. 

In Section~\ref{internal_tt} we investigated the relative accuracy of the intensity scale internal to the survey on the basis of total-intensity data, and reached the conclusion that the internal scale has a probable error of $2\%$.

In Section~\ref{angle_tests} we presented results for polarization angle in the
Fan Region. The scatter of measurements over 108 square degrees is
$8^{\circ}$\,rms (Table~\ref{fan_region_angles}). This value does not reflect
thermal noise; it incorporates, and is dominated by, the systematic and random
effects that influence the determination of angle. Although the sky directions
involved are close together, that does not mean that the observations were close
in time. In fact, our observation technique ensured that measurements  of
neighbouring points were well spread out in time, and the basketweaving process
brought a large number of observations, made over a long time period, to bear on
the determination of every data point. We therefore adopt $8^{\circ}$ as the
probable error in angle of our data. This includes a small contribution arising
from the fact that we have not corrected for ionospheric Faraday rotation (see
Section\,\ref{final}). To this error we must add $1.5^{\circ}$ for the
systematic error in our calibration of angle using the WMAP data.

The sky was not uniformly sampled by our observing technique. The observing scheme, described in Section~\ref{obs}, of half the scans terminating at declination $60^{\circ}$ and half terminating at $87^{\circ}$, was designed to spread the available observing time more optimally over the sky. Despite this, sampling at high declinations was still more thorough than sampling at low declinations, and we might expect greater sensitivity at high declinations. Nevertheless, it was difficult to discern any systematic improvement in survey data at high declinations, possibly because of systematic effects. We note that 
the principal product of the survey, the Faraday depth cube, is derived from angle data. As pointed out above, uncertainties in angle data are dominated by systematic errors, not by thermal noise.

\section{Results}
\label{results}

In this section we present a few results from the survey. We describe a check of
the quality of the Faraday cube, and show one example that illustrates some of
the scientific potential of the data. Table~\ref{fd_specs} gives details of the published data.

The spectral moments of the Faraday cube \citep{dick19} provide a very succinct portrait of multi-channel polarization data, and we use them here. The zero moment is the total polarized intensity integrated over the full range of $\phi$; it is defined as
\begin{equation}
{M_0}~~{\equiv}~~{{\sum^{n}_{i=1}}{T_i}\thinspace{\Delta\phi}},
\end{equation}
with units ${\rm{K}}\,{\rm{rad}}\thinspace{\rm{m}}^{-2}$, where $T_i$ is the polarized intensity at channel $i$, and $\Delta\phi$ is the width of each of the $n$ channels of the Faraday spectrum contributing to the sum. The first moment is the intensity-weighted mean of the Faraday depth. In Faraday simple directions this is the same as the peak Faraday depth. The first moment is
\begin{equation}
	{M_1}~~{\equiv}~~
{\frac{{\sum\limits^{n}_{i=1}}{T_i}{\phi_i}}{{\sum\limits^{n}_{i=1}}{T_i}}},
\end{equation}
with units ${\rm{rad}}\thinspace{\rm{m}}^{-2}$.  We excluded Faraday depth
channels beyond $\pm$500 ${\rm{rad}}\thinspace{\rm{m}}^{-2}$ in the moment
calculations to avoid contamination by spurious peaks (see
Section~\ref{rmsynth}), and excluded channels having polarized intensity below
0.04 K (0.01 K higher than the CLEAN threshold). For each pixel, Faraday depth
peaks with polarized intensities lower than 15\% of the primary peak in that
spectrum were also excluded.

Figure~\ref{I_moment0} shows Stokes $I$ at 1497\,MHz and the zero moment map.
Figure~\ref{angle_moment1} presents a map of polarization angle, $\chi$, at
1497\,MHz and first moments computed from the Faraday cube. Inspection of
Figure~\ref{angle_moment1} shows some areas where the Faraday depth is
significantly non-zero. This discovery was the basis of the first scientific
paper from this survey: \cite{woll10b} demonstrated the association of strong
features in Faraday depth with a large \ion{H}{1} bubble.

Close inspection of the images in Figures~\ref{I_moment0} and
\ref{angle_moment1} reveals some artefacts near the southern survey limit. These
arise from imperfect removal of ground emission and other instrumental effects.
The artefacts vary with frequency and position, and are very difficult to
quantify. They are confined within $5^{\circ}$ of the southern limit.  The sky
at the southern limit of the survey, declination $-30^{\circ}$, was observed at
an elevation of only $11^{\circ}$, above an uneven, mountainous horizon. We
anticipated that correction for ground radiation would be difficult under these
circumstances, but we chose this southern limit in order to maximize overlap
with GMIMS surveys of the southern sky (e.g. \citealp{woll19}). Readers should
exercise caution in using survey data between declinations $-25^{\circ}$ and
$-30^{\circ}$.

\begin{figure*}
\centering
\includegraphics[angle=0, width=1.2\columnwidth,clip=]
        {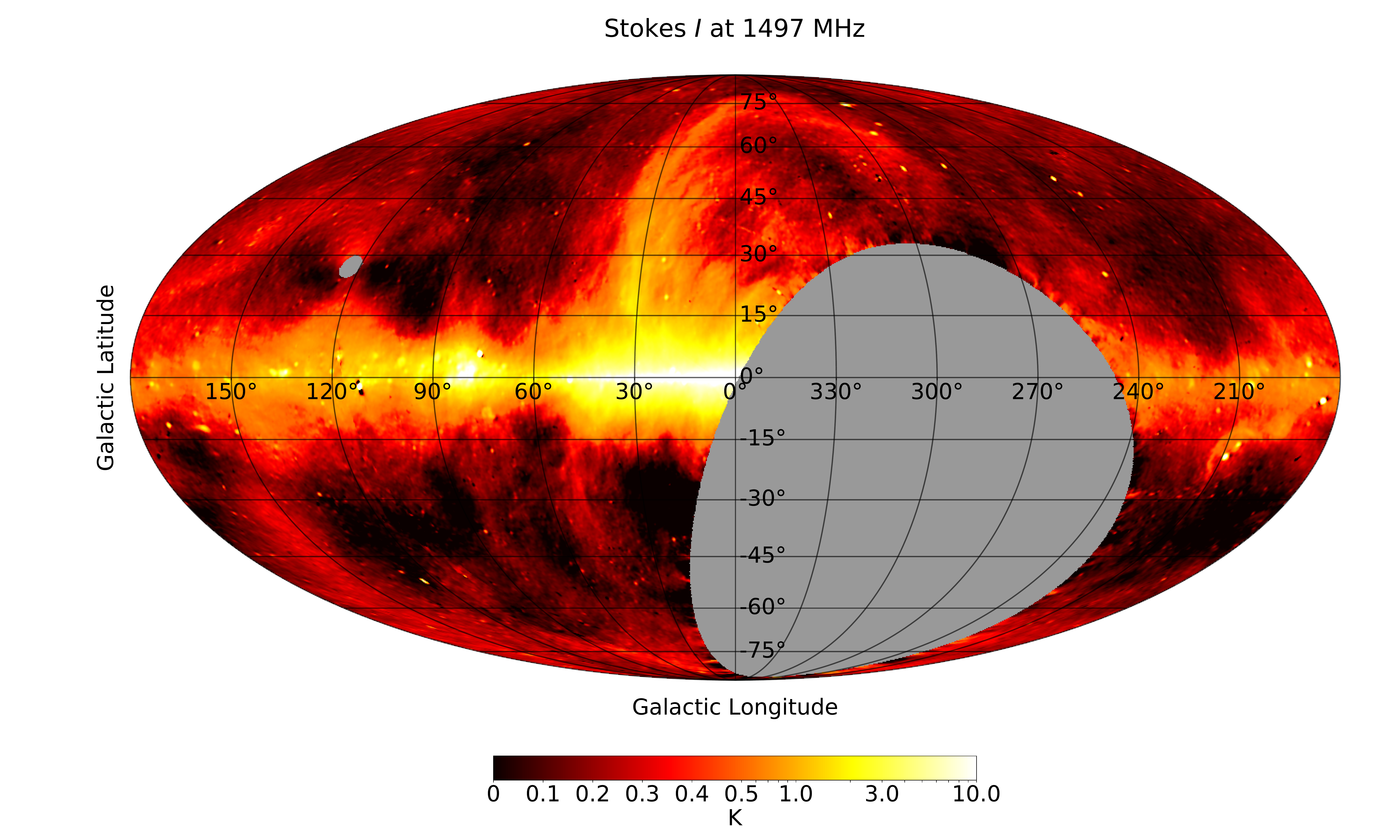}
	\includegraphics[angle=0, width=1.2\columnwidth,clip=]
        {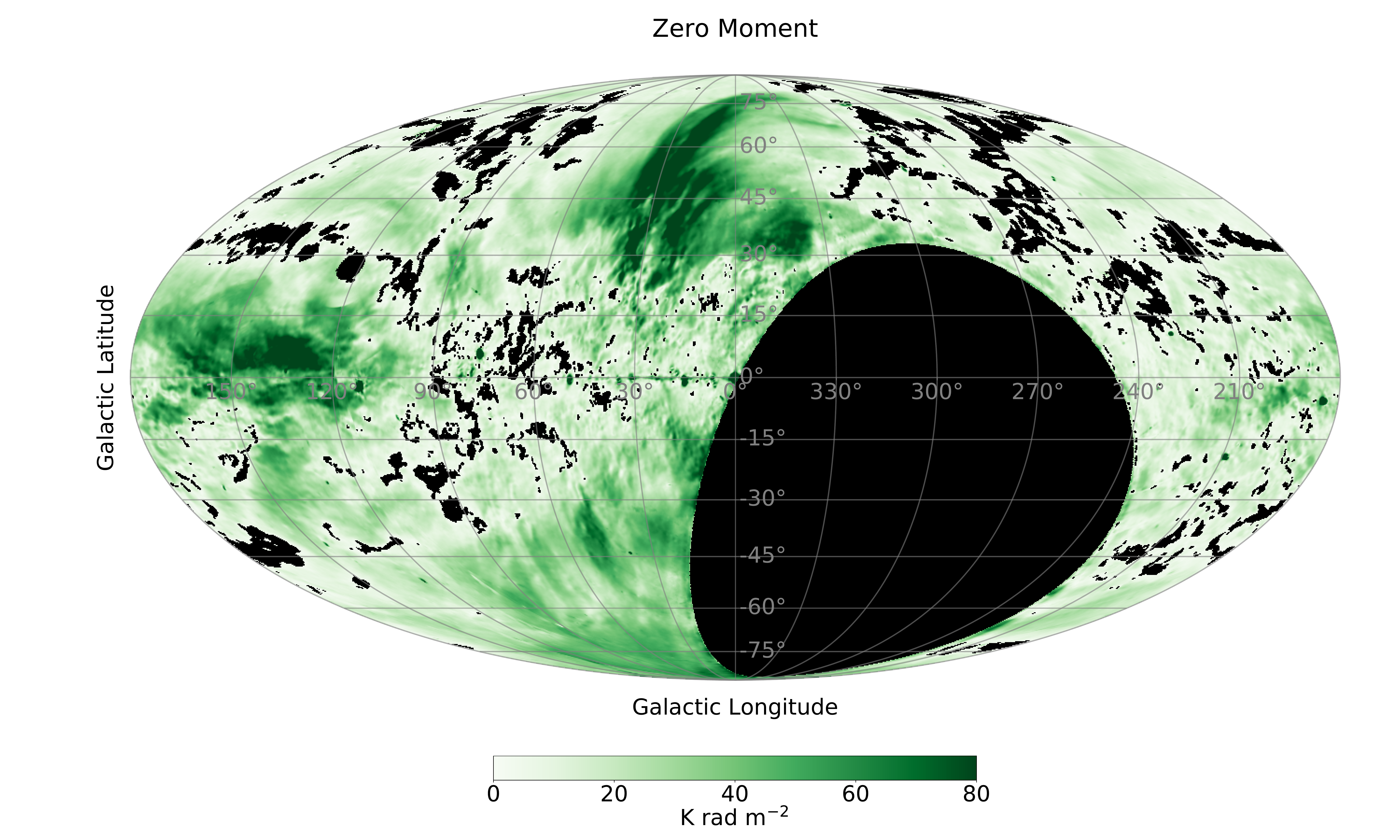}
 \caption{Top: total intensity at 1497\,MHz. The intensity scale of this image is correct, but the zero level is not (see text). Pixels outside the survey limits are gray. Bottom: zeroth moment computed from the Faraday cube, using the equations of \citet{dick19}. Pixels with insufficient data for the moment calculation, and those outside the survey area, are black. Both images are plotted in Galactic coordinates in Mollweide projection.}
 \label{I_moment0}
\end{figure*}

\begin{table*}
\caption[]{Characteristics of published survey data}
\label{fd_specs}
\begin{center}
\begin{tabular}{ll}
\hline
Frequency range, $I$, $Q$ and $U$ & 1280 to 1750\,MHz \\
Channel width                     & 1.1804\,MHz \\ 
Available data formats & Galactic coordinates, fits and healpix \\
Noise, $Q$ and $U$ images (single channel) & 45\,mK \\
Noise, $I$ images (50\,MHz band) & 20\,mK \\
Probable error, amplitude scale & 8\% \\
Probable relative error, internal intensity scale & $2\%$ \\
Probable error, polarization angles & $8^{\circ}$ \\
Systematic error, calibration of polarization angle & $1.5^{\circ}$ \\
Coverage of Faraday cube   & ${\pm}500$\,${\rm{rad}}\thinspace{\rm{m}}^{-2}$\\
Channel width in Faraday cube  & 5\,${\rm{rad}}\thinspace{\rm{m}}^{-2}$ \\
Largest Detectable Faraday depth & ${\sim}2{\times}10^4\,{\rm{rad}}\thinspace{\rm{m}}^{-2}$\\
Resolution in Faraday depth  & 150\,${\rm{rad}}\thinspace{\rm{m}}^{-2}$ \\
Largest measurable RM Structure & 110\,${\rm{rad}}\thinspace{\rm{m}}^{-2}$ \\ Sensitivity in Faraday Depth cube & 3.3 mK (rms) of polarized intensity \\
\hline
\hline
\end{tabular}
\end{center}
\end{table*}

\begin{figure*}
\centering
\includegraphics[angle=0, width=1.2\columnwidth,clip=]
        {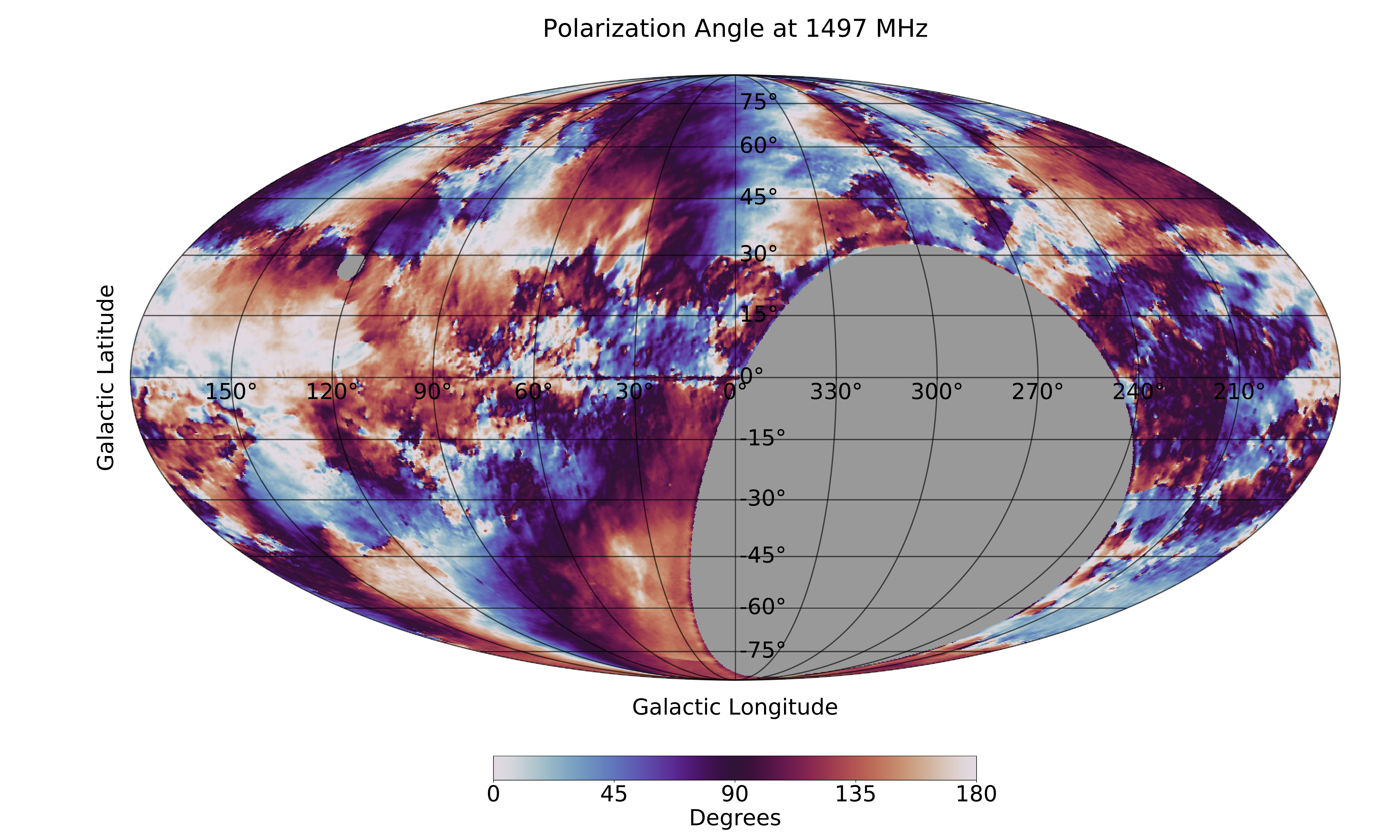}
\includegraphics[angle=0, width=1.2\columnwidth,clip=]
        {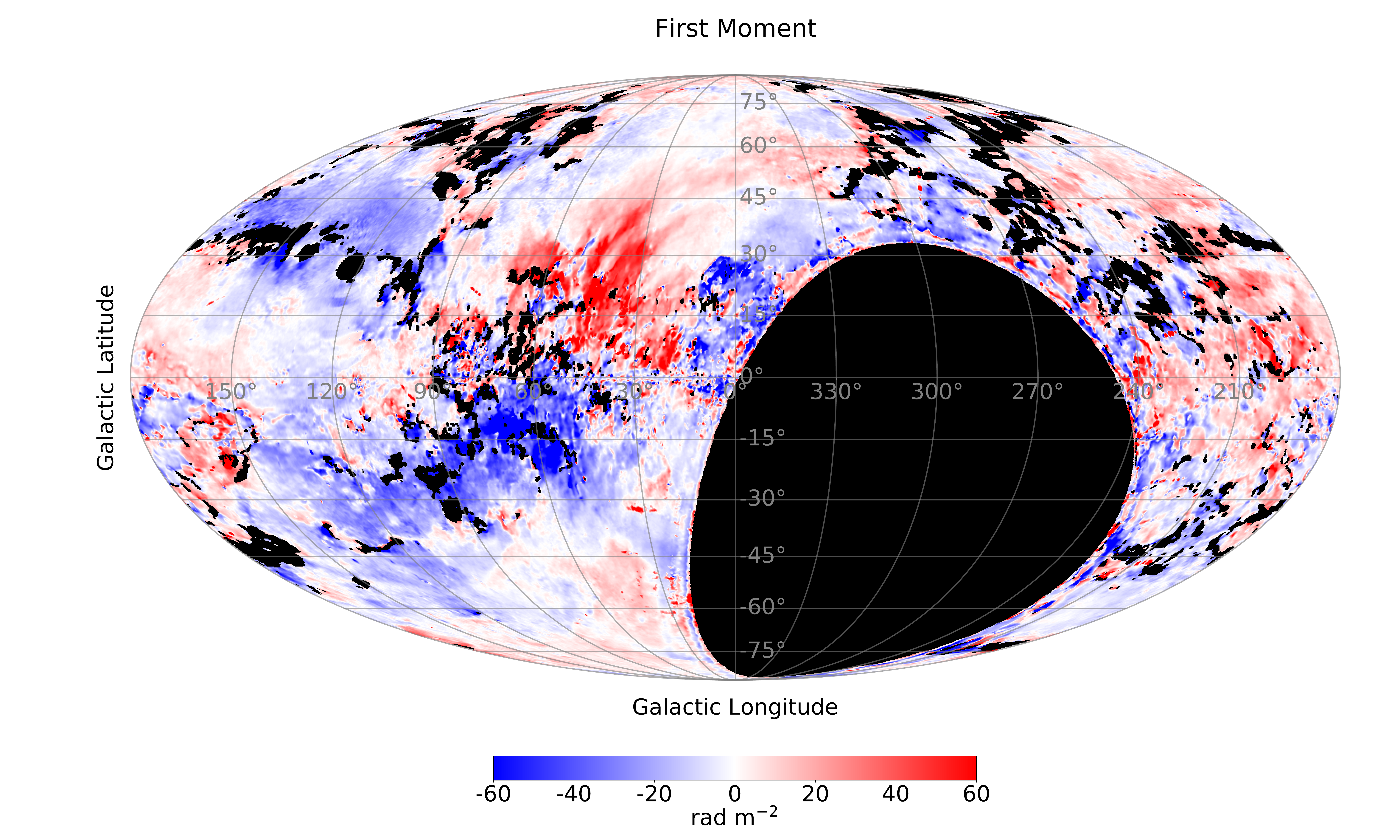}
 \caption{Top: polarization angle at $1497$\,MHz. Pixels outside the survey limits are gray. Bottom: first moment computed from the Faraday cube, using the equations of \citet{dick19}. Pixels with insufficient data for the moment calculation, and those outside the survey area, are black. Both images are plotted in Galactic coordinates in Mollweide projection.}
 \label{angle_moment1}
\end{figure*}

\begin{figure}
   \centering
   {\includegraphics[width=8.5cm,clip]{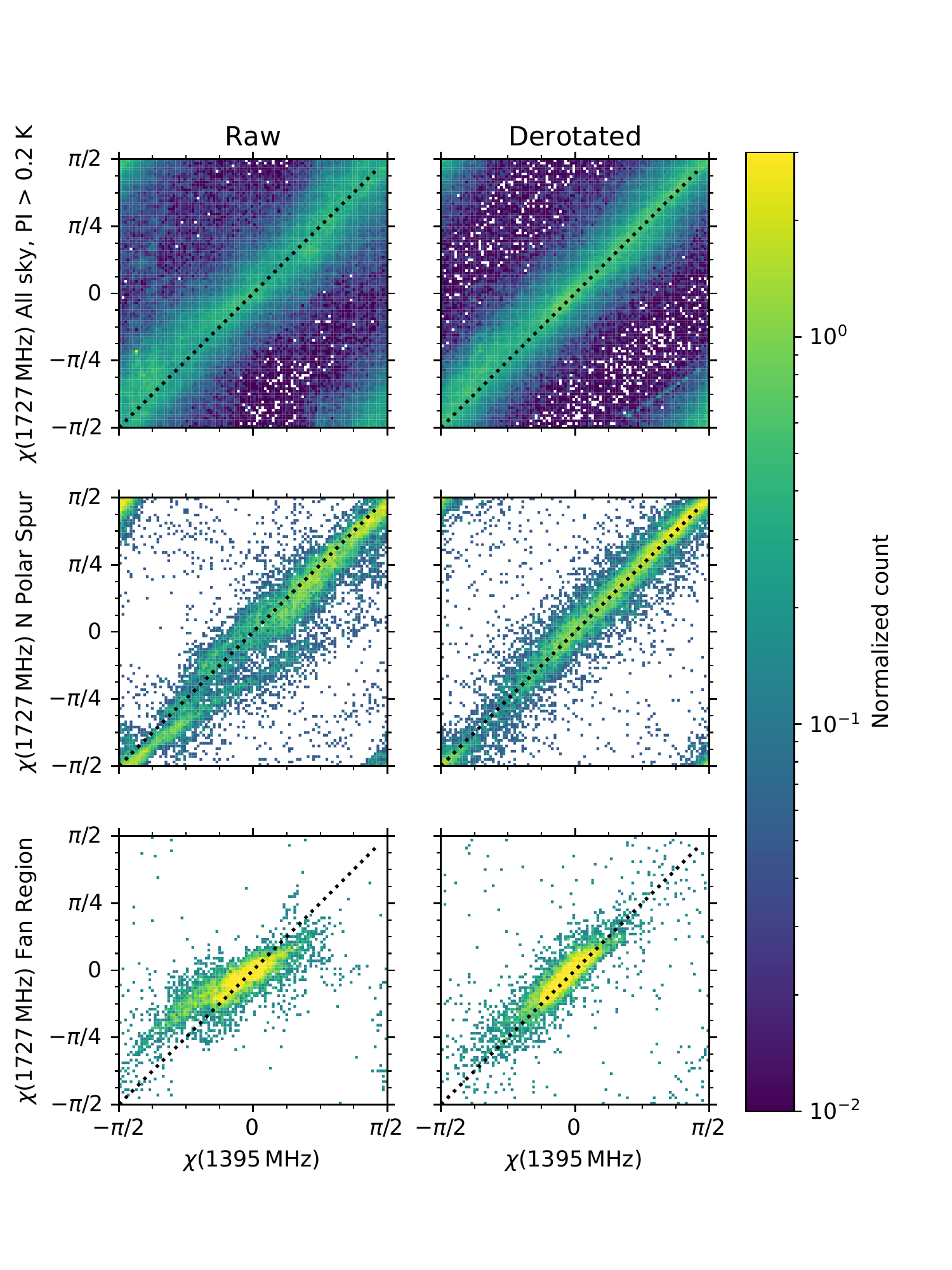}}
      \caption{Polarization angle at 1725\,MHz plotted point by point against polarization angle at 1394\,MHz. Top row: the entire survey, for points where the polarized intensity exceeds 0.2\thinspace{K}. Middle row: North Polar Spur. Bottom row: Fan Region. Left column, polarization angles as observed. Right column: the same scatter plot but the angles at 1394\,MHz and 1725\,MHz have been rotated by angle $-{\phi}\thinspace{\lambda^2}$ where $\phi$ is the Faraday depth deduced from the First Moment map.}
        \label{theta-theta}
\end{figure}

\begin{figure}
   \centering
   {\includegraphics[width=8.5cm,clip]{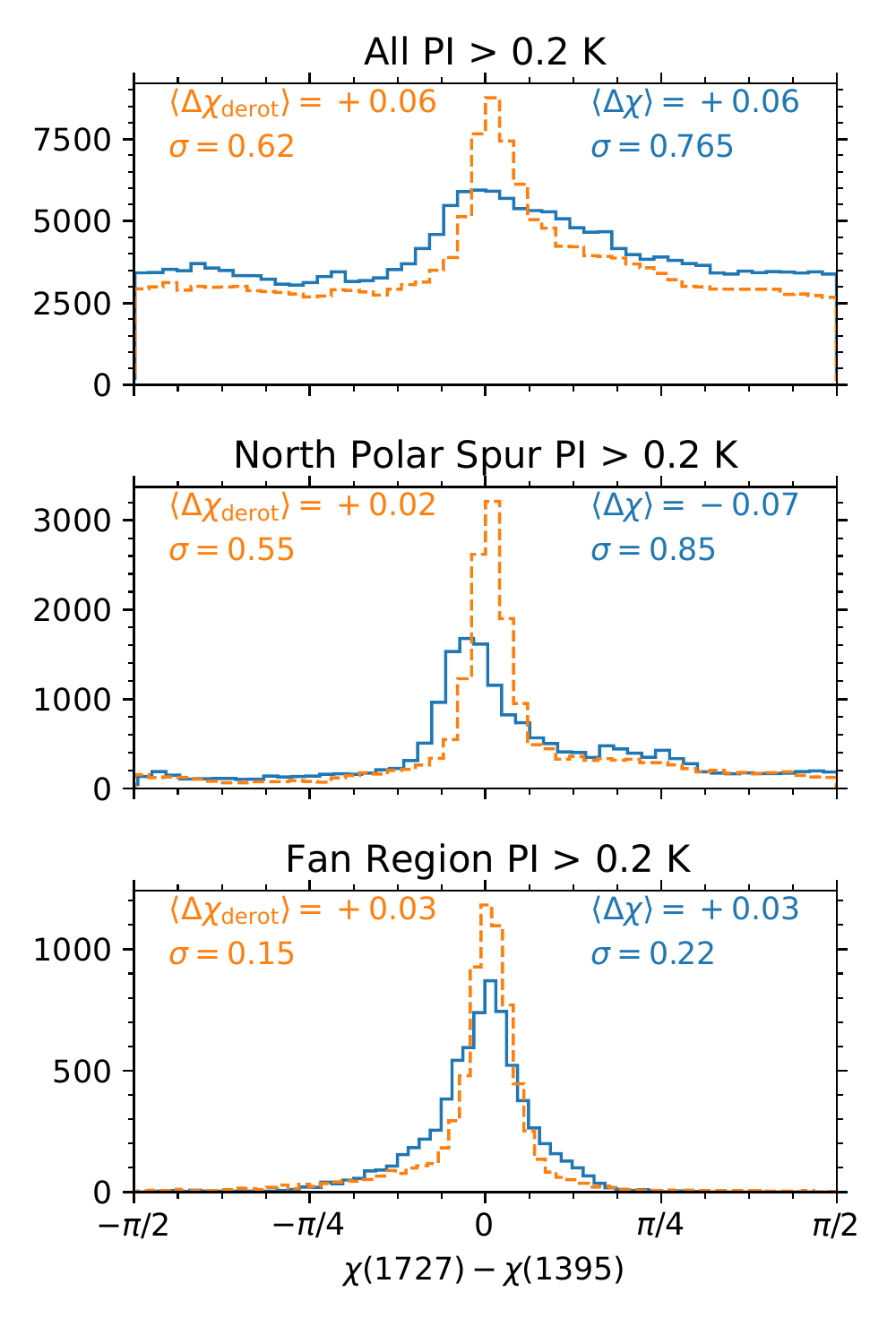}}
      \caption{Histograms of the differences in polarization angle for the scatter plots of Figure~\ref{theta-theta}. The mean, ${\langle}{\Delta}{\chi}{\rangle}$, and standard deviation, $\sigma$, for each histogram are indicated.}
   \label{theta-hist}
\end{figure}

\subsection{Quality of the Faraday cube}

In Figure~\ref{theta-theta} we present $\chi$-$\chi$ plots, comparing
polarization angles at two frequencies within this survey. The left column
compares observed polarization angles at 1394\,MHz and 1725\,MHz. In the right
column, we derotated the angles at both frequencies as
\begin{equation}
\chi_\mathrm{derot} = \chi_0 - \phi \lambda^2,
\end{equation}
where $\chi_0$ is the raw angle and $\phi$ is the moment-1 Faraday depth at each
pixel. The rows in Figure~\ref{theta-theta} contain the whole sky, the North
Polar Spur, and the Fan Region. In the raw $\chi$-$\chi$ plots, the agreement
between polarization angles is already good, as expected because there is
relatively little Faraday rotation, given the fairly small Faraday depths and
short wavelengths in this survey. However, there are notable deviations from the
$1:1$ line which are especially evident in the North Polar Spur and Fan Region.
In particular, there are a significant number of points at which the 1727\,MHz
angles are smaller by $\pi/8$ radians than the 1395\,MHz angles in the North
Polar Spur and at which the 1727\,MHz angles are larger by a comparable amount
than the 1395\,MHz angles in the Fan Region, especially near $\chi(1395\,MHz)
\approx -\pi/8$.

The derotation process brings most of the points for which the angles are discrepant from $1:1$ back in line. The histograms and associated statistics in Figure~\ref{theta-hist} show that the derotation process also reduces the scatter in all three samples: the distribution of derotated polarization angles is more centrally peaked and narrower than the distribution of raw polarization angles, and the standard deviations of the distributions are reduced by $18-35\%$. Moreover, in both Figure~\ref{theta-theta} and Figure~\ref{theta-hist} it is evident that the derotated data have fewer points in which the angles differ by $\sim \pi/4$ or, equivalently, are far from the 1:1 lines in Figure~\ref{theta-theta}. We interpret this as an indication of the Faraday simplicity of the data in this band as well as a check on the efficacy of the RM synthesis procedure.

If the observed Faraday rotation were idealized such that $\chi$ was a straight line as a function of $\lambda^2$, we would expect this derotation process to produce perfect agreement across frequencies; in this case, the Faraday synthesis process would have been unnecessary in the first place, and we could have simply measured $\mathrm{RM} = d \chi / d \lambda^2$. We do not observe this: there is noticeable scatter about the 1:1 line. This is not surprising: it is simply a confirmation that the interstellar medium is not Faraday simple. We take the tight relationship of polarization angle across the band, in particular after derotating, as a check on the internal consistency of the polarization angle measurements in this survey.

\begin{figure*}[!h]
   \centering
   {\includegraphics[width=8cm,clip]{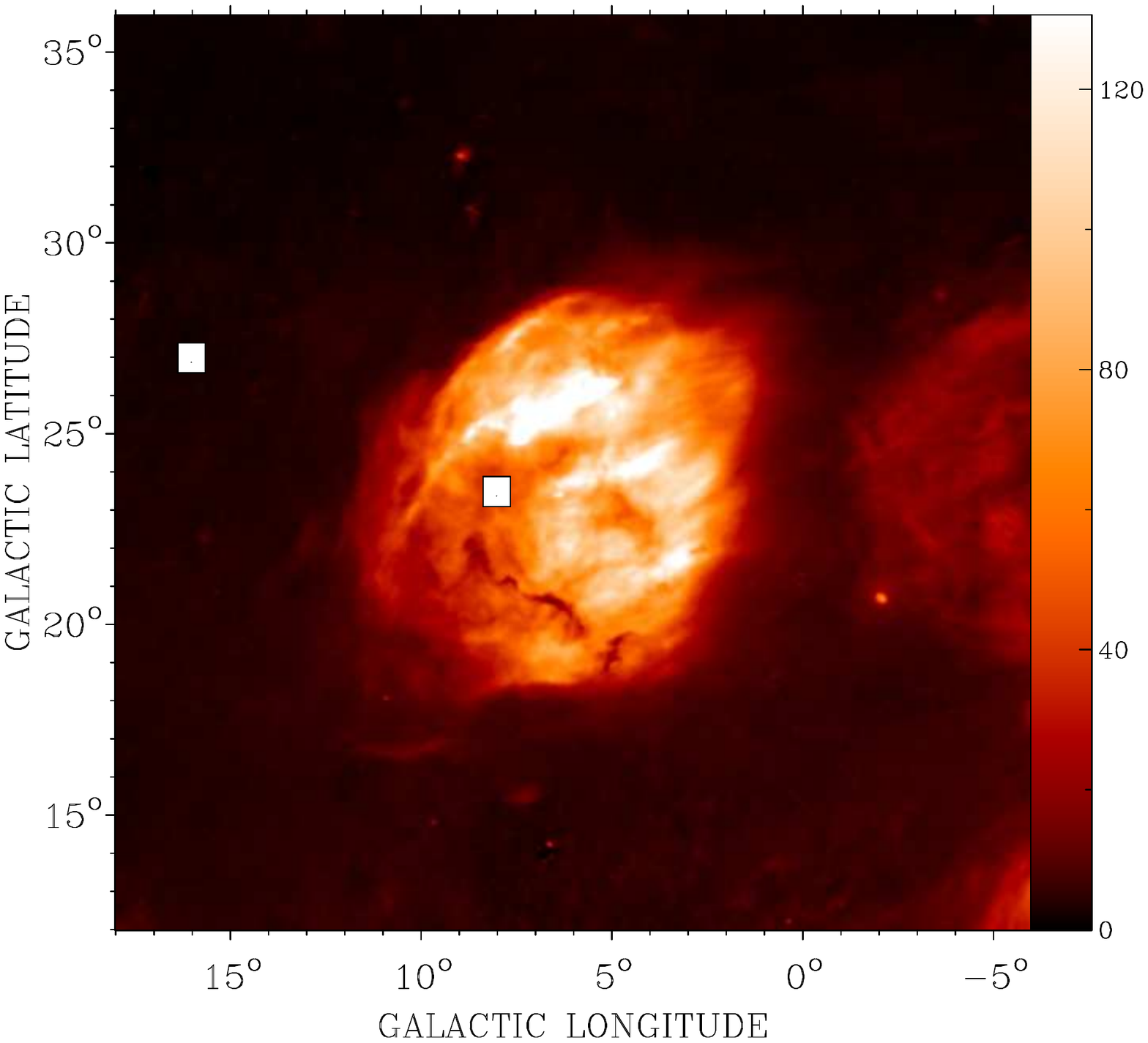}}
   {\includegraphics[width=8cm,clip]{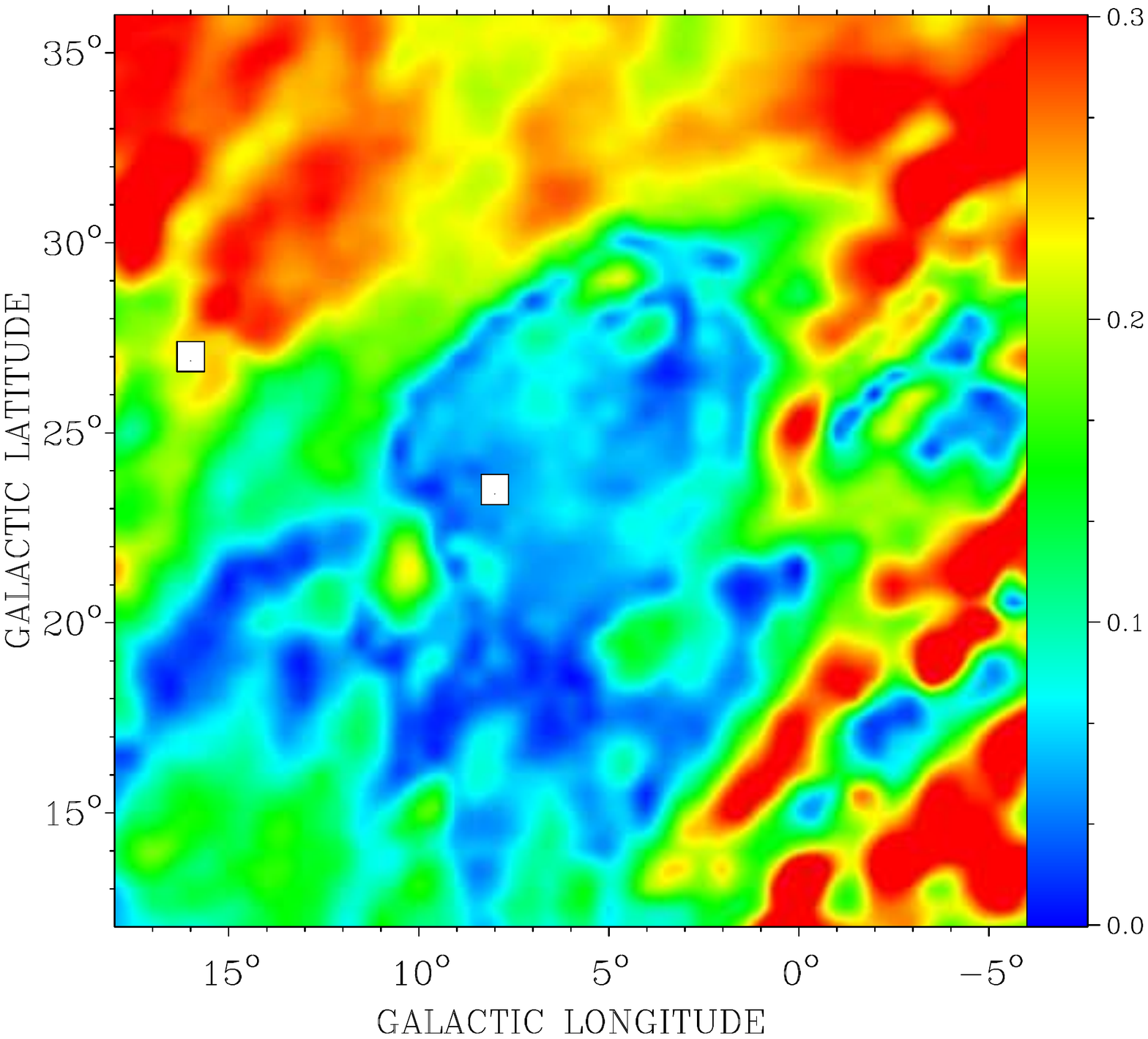}}  
   \caption{The \ion{H}{2} region Sharpless 2-27. The left image shows H$\alpha$ image in units of Rayleighs from the data of \citet{fink03}. The right image shows polarized intensity at a Faraday depth of $-$55\,${\rm{rad}}\thinspace{\rm{m}}^{-2}$ in units of K\,RMSF$^{-1}$. Two white squares are superimposed on these images: these are the locations of the two Faraday spectra shown in Figure~\ref{s27_spectra}.}
   \label{sharpless}
\end{figure*}

\begin{figure*}[!h]
   \centerline{\includegraphics[bb = 1 1 1100 400,width=16cm,clip]
   {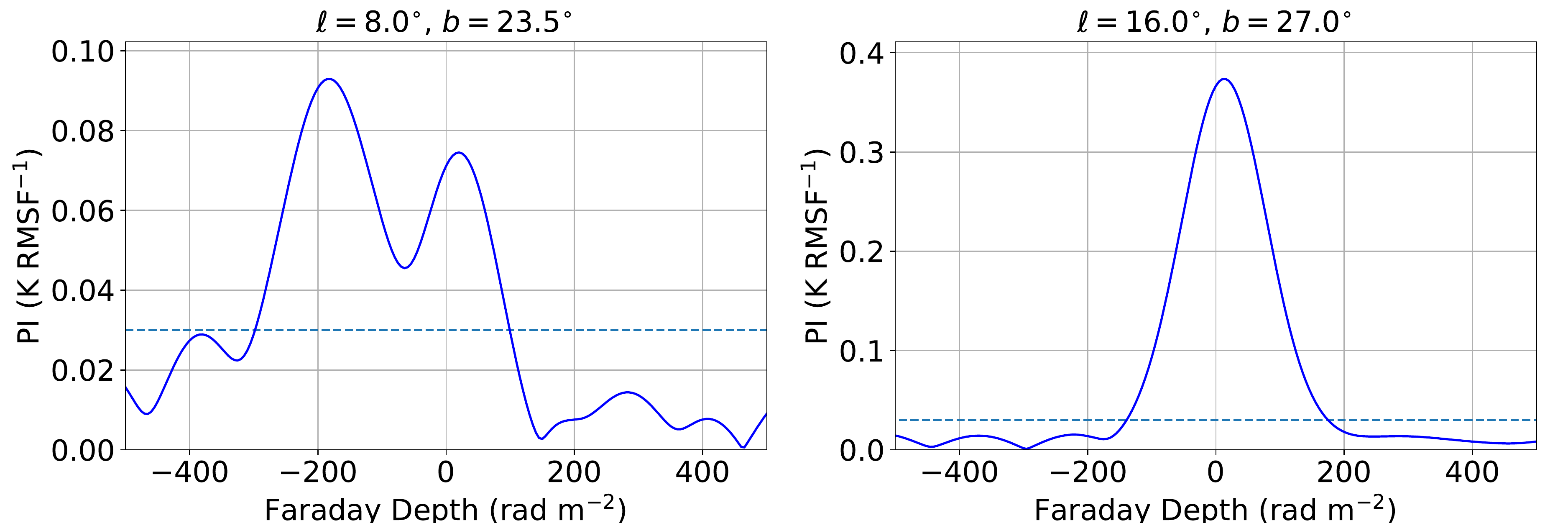}}
   \caption{Faraday spectra at ${(\ell,b)}= {(8^{\circ},23.5^{\circ})}$ and
   $(16^{\circ},27^{\circ})$. The dotted line on each plot shows the clean limit, 0.03\,K\,RMSF$^{-1}$.}
   \label{s27_spectra}
\end{figure*}

\subsection{The \ion{H}{2} Region Sharpless 2-27}

As an example of insights to be gained from the Faraday depth cube, we discuss FD spectra in the direction of the \ion{H}{2} region Sharpless 2-27 (which we refer to as S27). S27, at ${(\ell,b)}= {(6.3^{\circ},23.6^{\circ})}$ is a large \ion{H}{2} region, ${\sim}10^{\circ}$ in extent, excited by the star $\zeta$Oph whose distance is 180\,pc \citep{vanl08}. Figure~\ref{sharpless} shows the object in H$\alpha$ from the data of \citet{fink03}, and in our data at a Faraday depth of $-$55\,${\rm{rad}}\thinspace{\rm{m}}^{-2}$. Figure~\ref{s27_spectra} shows the Faraday depth spectra at two points, one on S27 and one in a nearby direction off the \ion{H}{2} region. 

\citet{thom19} have analyzed data from the GMIMS 300--480\,MHz survey \citep{woll19} in the direction of S27. At those low frequencies, the \ion{H}{2} region totally depolarizes background emission and the Faraday spectrum reveals details of the synchrotron emission generated in the foreground column. It is evident from Figure~\ref{s27_spectra} that S27 has strong Faraday effects in the 1280-1750\,MHz band as well. Here we present only an outline interpretation of the data. We will present a full analysis of our data in this direction in a forthcoming paper (Ordog et al, in preparation). 

In the frequency range of the present work, background emission is strongly depolarized and Faraday rotated by S27. The off-source FD spectrum in Figure~\ref{s27_spectra} peaks at a polarized intensity of 0.37\,K\,RMSF$^{-1}$ and there is only one peak, at about +15\,${\rm{rad}}\thinspace{\rm{m}}^{-2}$. The on-source spectrum shows two peaks, at +22 and $-$190\,${\rm{rad}}\thinspace{\rm{m}}^{-2}$, and polarized intensity reaches no higher than 0.09\,K\,RMSF$^{-1}$. We interpret this spectrum as showing foreground emission from the 165\,pc path{\footnote{The distance, 180\,pc, to the exciting star minus the 15\,pc radius of S27.}} between S27 and the telescope (the peak at positive Faraday depth), and background emission Faraday rotated on passing through S27 (the peak at negative Faraday depth). Faraday rotation through S27 was identified by \citet{harv11} by examination of RMs of background sources seen through the object using RMs from the catalog of \citet{tayl09}. The two sources from that catalog closest to ${(\ell,b)}={(8^{\circ},23.5^{\circ})}$ have an average RM of $-$217\,${\rm{rad}}\thinspace{\rm{m}}^{-2}$. The average RM of six sources within a $3^{\circ}$ circle around that position is $-$162\,${\rm{rad}}\thinspace{\rm{m}}^{-2}$. These values compare well with the Faraday depth at that position in our data.

\section{Conclusions}
\label{concl}

We have described observations and data processing which have yielded Stokes
parameters $I$, $Q$, and $U$ over the Northern sky, between declination limits
of $-30^{\circ}$ and $+87^{\circ}$, covering 72\% of the sky; 95\% of full
Nyquist sampling has been achieved. Frequency coverage is 1280 to 1750\,MHz.
Although much of this frequency band lies outside the ranges allocated to radio
astronomy, the data loss to RFI is only $\sim$30\%. This work was designed as a
Faraday depth survey, not simply a polarization survey, and its most valuable
published data product is a Faraday depth cube, covering
${\pm}$500\,${\rm{rad}}\thinspace{\rm{m}}^{-2}$. We have achieved a sensitivity
of 3\,mK and a resolution in Faraday depth of
150\,${\rm{rad}}\thinspace{\rm{m}}^{-2}$. However, our sensitivity to wide
structures in Faraday depth extends only as far as
110\,${\rm{rad}}\thinspace{\rm{m}}^{-2}$. Future plans for the GMIMS
project include observations at lower frequencies. When coverage is extended
down to 800\,MHz the resolution in Faraday depth will improve to
$\sim$35\,${\rm{rad}}\thinspace{\rm{m}}^{-2}$ with the same sensitivity to
extended FD structures. We will then be able to identify wide structures in Faraday depth without ambiguity.

Users of the data should be aware that we have concentrated on an accurate
depiction of the extended polarized emission. Furthermore, our observing
technique is not ideal for the measurement of compact sources, and data on such
sources extracted from our survey should be treated with caution.  We note,
again, that basketweaving has removed the sky minimum from total-intensity
images; any use of the total-intensity data must take this fact into account.

The survey has been calibrated against absolute standards of noise and the
well-established flux density and spectrum of Cygnus\,A, and all data products
are in units of absolutely calibrated main-beam brightness temperatures. This
was necessary because no comparable surveys were available as calibrators,
except near 1400\,MHz. All GMIMS surveys are (or will be) absolutely calibrated,
and this allows accurate intercomparison and combination of data from different
component surveys. Comparison with available total-intensity data near 1400\,MHz
demonstrates very satisfactory agreement of scales, within 5\%. Comparison with
available polarization data indicates agreement of the polarized intensity scale
within 10\%.  The intensity scale within the 1280 to 1750 MHz passband is
consistent within 2\%. We have demonstrated a new technique for calibration of
polarization angle using lines of sight in the Fan Region which we have
identified as having zero Faraday rotation; we have calibrated our data using
the WMAP 23\,GHz dataset. This technique can be applied to any polarization
survey in the North, and will be used with future GMIMS surveys.

We encountered some difficulty with estimation of the ground contribution. In that regard our technique of making observations by moving the telescope in elevation is not ideal. The technique developed by \cite{carr19} for the SPASS survey, scanning in azimuth, is superior, but the equatorial mounting of the Galt Telesope ruled out that as a possibility.

This is the second GMIMS survey to be published, following the 300--480\,MHz
survey of the Southern sky with the Parkes 64-m Telescope \citep{woll19}. The
overlap between the surveys spans the declination range $-30^{\circ}$ to
$+20^{\circ}$, 42\% of the sky. This overlap has already been exploited by
\citet{dick19} to compare the two surveys, and it has great future potential.
Once again we find that a large fraction of the sky displays significant
polarized emission at non-zero Faraday depth. This was not apparent from
polarization surveys at single frequencies, of course, and provides a rich
opportunity for investigation.

The data presented here are available at the Canadian Astronomy Data Centre, at http://dx.doi.org/10.11570/21.0003.

\acknowledgments

We are indebted to Rob Messing for his skill in designing, building, and
maintaining the receiver through the course of this project. We thank Xuan Du
and Tim Robishaw for outstanding contributions to the determination of aperture
efficiency of the telescope. Stasi Baran helped greatly with the observations.
We thank Sean Dougherty for his support of this project. The Dominion Radio
Astrophysical Observatory is a National Facility operated by the National
Research Council Canada. Janet Weiland of Johns Hopkins University provided us
with information on systematic errors in WMAP data. The Global Magneto-Ionic
Medium Survey is a Canadian project with international partners. The
participation of MW, KHD, AO and ASH was supported in part by the Natural
Sciences and Engineering Research Council (NSERC). The work of ASH and AO was
also supported by the Dunlap Institute and the National Research Council Canada.
The Dunlap Institute is funded through an endowment established by the David
Dunlap family and the University of Toronto. BMG acknowledges support from NSERC
and the Canada Research Chairs program. JLH is supported by the National Natural
Science Foundation of China (NNSFC grants 11988101 and 11833009) and the Key
Research Program of the Chinese Academy of Sciences (grant QYZDJ-SSW-SLH021).
This research has been enabled by the use of computing resources provided by
WestGrid and Compute/Calcul Canada and the Centre for High Performance Computing
in Cape Town, South Africa.

\bibliographystyle{aa}
\bibliography{survey_arxiv}

\end{document}